\documentclass[%
 reprint,
 twocolumn,
 amsmath,amssymb,
 aps,
]{revtex4-2}
\usepackage[utf8x]{inputenc}
\usepackage[english]{babel}
\usepackage{graphicx}
\usepackage{dcolumn}
\usepackage{bm}
\usepackage{float}
\usepackage{amsfonts} 
\usepackage{tikz}
\usepackage{pgfplots}
\usepackage{multirow}
\usepackage{circuitikz}
\usepackage{boldline}
\usepackage{comment}
\usepgfplotslibrary{fillbetween}
\usetikzlibrary{calc}
\usetikzlibrary{arrows}
\usetikzlibrary{patterns}
\usetikzlibrary{fadings}
\usetikzlibrary{arrows.meta}
\usetikzlibrary{topaths}
\usepackage[skip=5pt, singlelinecheck=off, justification=raggedright]{caption}
\usepackage{subcaption}
\usepgfplotslibrary{groupplots}
\tikzfading[name=downfading, bottom color=transparent!100, top color=transparent!50]
\tikzfading[name=upfading, bottom color=transparent!50, top color=transparent!100]

\usepackage[compat=1.1.0]{tikz-feynman}

\newcommand{\bs}{\boldsymbol}

\DeclareMathSymbol{\shortminus}{\mathbin}{AMSa}{"39}

\begin{document}

\preprint{APS/123-QED}

\title{Bose-Einstein condensation of THz photons in an optical microcavity with Landau-quantized electrons}

\author{Timofey V. Maximov}
\email{TVMaksimov@vniia.ru}
\affiliation{N.L. Dukhov Research Institute of Automatics (VNIIA), Moscow 127030, Russia}

\author{Norayr A. Asriyan}
\affiliation{N.L. Dukhov Research Institute of Automatics (VNIIA), Moscow 127030, Russia}

\author{Igor L. Kurbakov}
\affiliation{Institute for Spectroscopy RAS, Troitsk 142190, Moscow, Russia}

\author{Yurii E. Lozovik}
\affiliation{Institute for Spectroscopy RAS, Troitsk 142190, Moscow, Russia}


\begin{abstract}

We present a theoretical model for a coherent terahertz radiation source based on Bose-Einstein condensate of incoherently pumped microcavity photons. Energy relaxation is provided by inelastic photon scattering on a two-dimensional electron gas in magnetic field. The proposed setup evades the standard lasing mechanisms: neither population inversion nor light wave amplification is utilized. We study the kinetics of photon condensation and describe a semiconductor-crystal based device.
\end{abstract}

\maketitle


\section{Introduction}\label{sec:intro}

When decreasing the temperature in a system of Bose particles, the associated thermal de-Broglie wavelength increases and, eventually, exceeds the mean interparticle distance. That is where quantum effects manifest themselves, leading to Bose-Einstein condensation (BEC) as observed experimentally for cold atomic gases \cite{Anderson95, Davis95, Bradley97}. Further progress in experimental techniques led to observation of condensation signatures in quasiparticle systems such as excitons \cite{High12, Fogler14}, magnons~\cite{Demokritov2006}, excitonic polaritons \cite{Kasprzak09, Balili07} and microcavity photons \cite{Klaers10}. Due to finite quasiparticle lifetime, these are inherently open driven-dissipative systems, which encourages using them as a playground for investigating non-equilibrium phase transitions, critical behaviour, \textit{etc}. In addition, polaritonic/photonic condensates are of special interest due to facilitated access to the optical emission, which readily implies possible applications as coherent laser-type sources \cite{Yan00}.

For the occupation build-up in a single mode in these quasiparticle systems, one has to provide a source of enhanced influx to compensate the losses or minimize these losses directly. Typically, particles in the condensate are generated through scattering from high-energy states (as observed in exciton, exciton-polariton, and magnon condensates) or with the assistance of a gain medium. Our focus in this paper is on the photonic condensate in an optical microcavity. In these setups, the relaxation of the excited gain medium is routinely employed, given the negligible photon interaction.  Experimental studies have employed dye molecules \cite{Klaers10,Greveling2018,Walker2018} or semiconductor layers in VCSELs \cite{Schofield2024,Pieczarka2024,Fainstein2024} as the gain medium. The kinetics of condensate interaction with the gain medium in these experimental setups has been also investigated extensively~\cite{KirtonKeeling,Stoofdye, Radonjic2018, PhysRevA.100.043803, PhysRevLett.133.223601, PhysRevA.104.063709, Semicond_BEC_lasing} (see also the review \cite{Bloch2022}).

The condensate frequency in the mentioned experiments is in visible/IR range. Generating condensate of THz frequency in the same manner is not straightforward. Besides the need to find a medium with low excitation energy, it would require reduced temperatures (note, though, that coherent THz generation was realized with a coupled-cavity scheme with infrared VCSELs~\cite{HuBrenner}).

We propose and analyze an experimental scheme, which has a potential to overcome these difficulties. In terms of the geometry it is a typical VSCEL-type device, although the operation mechanism is different. Namely, we consider a photonic microcavity with an n-doped semiconductor layer with its electrons quantized by intense magnetic field. This setup allows us to suppress losses from the cavity's condensate mode via the emission of lattice photons. Additionally, our design provides conditions for the efficient relaxation of high-energy photons, injected by a thermal light source, into the condensate mode. In a sense, we present a means to adapt the kinetic condensation mechanism (typically observed for excitons and exciton-polaritons) for non-interacting photons within the microcavity. Thus, our theoretical analysis is qualitatively different from that present for existing experimental setups for photonic condensation.

A reliable THz source of the type described above would be of high demand for the needs of THz spectroscopy with a wide range of possible applications \cite{Lewis14,Zhang17}. THz radiation wavelength is in the range of 3 millimeters to 30 micrometers (frequency range is 0.1–10 THz), and so it possesses properties similar to those of long-wavelength radiation, being able to penetrate through non-conductive materials such as clothing, wood, plastic, ceramics \cite{Coutaz21}, thin layers of organic matter, to mention a few. THz photons are low-energy photons which do not destroy molecules they interact with. Experiments show that living tissue can withstand radiation power of about a dozen of milliwatts. That is the reason why THz radiation is a candidate for replacing highly energetic X-rays in various medicine \cite{Cherkasova16}, biochemistry \cite{Nazarov08}, non-destructive testing in studies of cultural and historical objects \cite{Dhillon17}, \textit{etc}.

It is also worth noting that many organic molecules have an oscillatory-rotational spectrum in the THz range, which makes it possible to use THz spectroscopy both to determine the spectral trace of certain molecules (electronic noise) and to excite specified vibrations in molecules. In addition to molecules, many collective vibrations in metals and semiconductors also possess THz spectra. The availability of an easy-to-use THz coherent radiation source would allow important experiments in the field of condensed matter as well \cite{Salen19,Valusis21}.

Currently, one has a bunch of techniques to choose from for THz generation. Namely, free electron lasers, quantum cascade lasers, optical laser based nonlinear converters, \textit{etc.} \cite{Lewis14, Dhillon17, Leitenstorfer23}. We claim that microcavity photonic condensate also belongs to this list, being a decent radiation source.

The article is organized as follows. We open Sec. \ref{sec:idea} with a presentation of the underlying principle of THz photon condensation and a schematic of the coherent light source, which is followed by a microscopic model. Then follows the core part of the paper: in Sec. \ref{sec:kinetics} we study the details of kinetic processes which feed and deplete the photonic condensate. The subsequent Sec. \ref{sec:analysis} is devoted to assessing and optimizing device performance by properly choosing control parameters (magnetic field, cavity Q factor, \textit{etc.}) in terms of the output power and efficiency. In addition, we address several technical issues of creating a device for practical usage. Sec. \ref{sec:conclusion} summarizes the results of our study. Finally, Appendices A-F present calculation details which, for the sake of clarity, are omitted in the main text.

\section{The THz coherent source}\label{sec:idea}

\subsection{The underlying principle}
\begin{figure}[htp]
    \begin{center}
    \includegraphics[width = 0.6\linewidth]{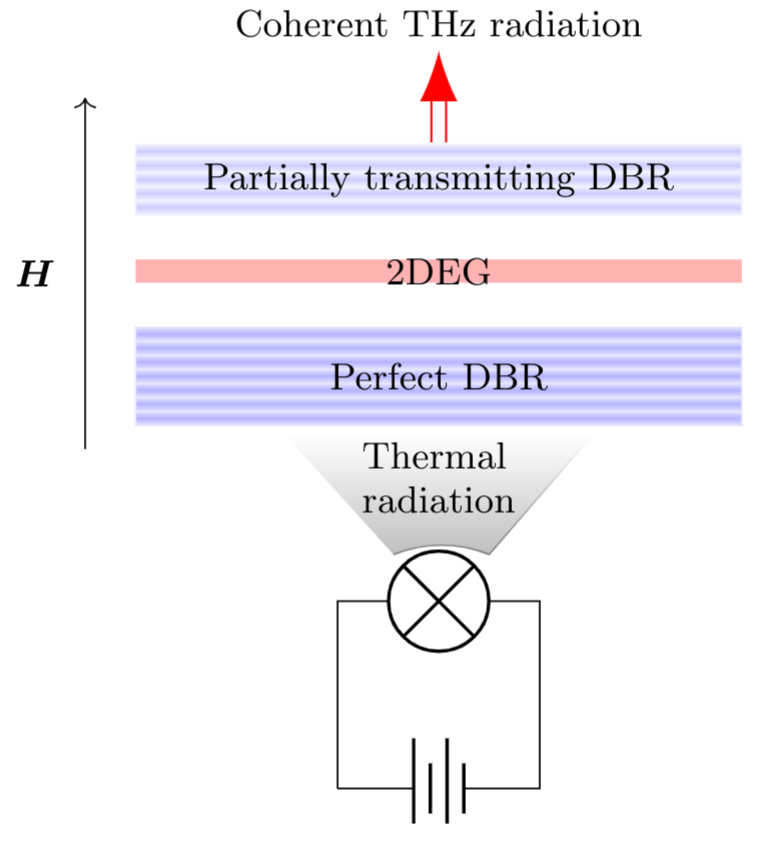}
    \end{center}
    \caption{Schematic of the coherent THz light source device. A layer with 2DEG is placed into an optical microcavity. External magnetic field is to quantize electron motion.}
    \label{fig:scheme}
\end{figure}

The scheme of the proposed device is presented in Fig. \ref{fig:scheme}. The aim is to achieve macroscopic population of a single photonic mode in an optical microcavity, illuminated by a thermal light source, in order for the emission for this mode outside the cavity to be coherent.

\begin{figure}[htp]
\begin{center}
    \includegraphics[width = 0.9\linewidth]{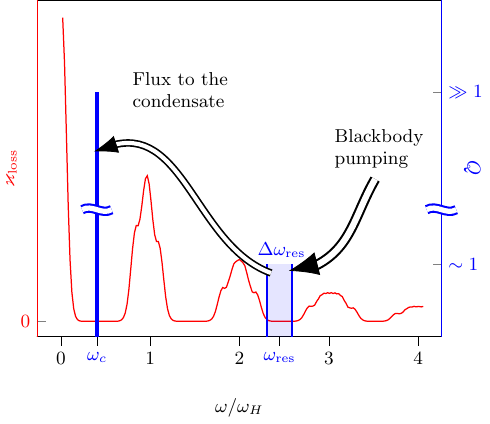}
    \end{center}
    \caption{Typical dependence of the photon absorption rate by the 2DEG $\varkappa_{\rm loss}$ as a function of photonic mode frequency (left axis, red dashed graph). The high-$Q$ condensate mode along with a bunch of low $Q$ pumping modes are depicted in blue.}
    \label{fig:cyclotron_idea}
\end{figure}

The main challenge in designing a device of the type described above is the dissipation. Photonic modes of the microcavity constitute an open system constantly exposed to interaction with the lattice and external pumping source. Due to the former, photons transmit the absorbed energy to the semiconductor, thus equilibrating with the lattice at relatively low temperatures. For an occupation buildup in a single mode, one has to increase the pumping rate of that mode and to suppress the leakage out of it. We propose using electron cyclotron resonance (ECR) to control the photon absorption rate by the 2DEG. As will be carefully demonstrated further, for a 2DEG in magnetic field, the rate of photonic outflow from a single optical mode of the microcavity (due to absorption by the electron gas) depends on its frequency in a periodic manner. This effect has been experimentally investigated for heterostructures in THz range (\textit{e.g.} see \cite{Cao18}). As schematically depicted in Fig. \ref{fig:cyclotron_idea}, peak values are close to the resonant absorption frequencies (the shapes of the peaks are due to the details of electron-phonon interaction). In contrast, the modes in the regions of low leakage are effectively isolated from the lattice.

With that in mind, one may engineer a cavity with a single mode of high Q-factor (with respect to the leakage through the mirrors) having a frequency in the minimum of the periodic profile. We will further refer to this mode as "condensate mode"\ (with frequency $\omega_c$ in Fig. \ref{fig:cyclotron_idea}).

However, low transmittance of the mirrors also prevents the essential efficient pumping of the condensate mode by the black body. To overcome this obstacle, we note that quantized electrons well-mediate interaction between photons in modes spaced by twice the cyclotron frequency $\omega_H$ (as will be demonstrated further, see \eqref{eq:nu-pump}). Having some photonic modes with a small $Q$-factor (thus, penetrable for external pumping) in the next, but one minimum of the leakage rate profile provides the necessary source of influx to the condensate mode. In addition, to broaden the frequency range $\Delta \omega_{\rm res}$ of these modes, as shown in Fig. \ref{fig:cyclotron_idea}, we propose introducing a magnetic field gradient over the semiconductor sample. We will further refer to these modes collectively as "pumping reservoir".

Note that, besides the condensate mode with zero in-plane momentum, there are multiple modes of the same energy (hereafter "satellite modes", see Fig. \ref{fig:kinetic_scheme}), which are also well-coupled to the condensate mode via elastic photon scattering on electrons.

Altogether, the pumping reservoir along with the condensate mode with its satellites make up a photonic subsystem, which is effectively isolated from both the lattice and the other photonic modes and exposed to black body pumping and dissipation through the Bragg mirrors. The kinetic processes in this subsystem are of interest in this work.

\subsection{Microscopic model}

To model the described device microscopically, we employ the following Hamiltonian, which characterizes all essential subsystems and their interactions:
\begin{align}
    &\hat{H} = \underbrace{\sum_{\bs{q},\xi}\hbar\omega_{\bs{q}}\hat{c}^\dagger_{\bs{q},\xi}\hat{c}_{\bs{q},\xi}}_{\rm microcavity\; photons} + \underbrace{\sum_{\bs{k},\zeta}E_{\rm ph}({\bs k},\zeta)\hat{b}^\dagger_{\bs{k},\zeta}\hat{b}_{\bs{k},\zeta}}_{\rm acoustic\; and\; optical\; phonons} \nonumber\\&
    + \underbrace{\sum_{\bs p,s}E_{\bs p}^s\hat{a}_{\bs p,s}^\dagger\hat{a}_{\bs p,s}}_{{\substack{\text{Landau-quantized} \\ \text{electrons}}}} \!\!{+}\underbrace{\sum_{\substack{{\bs p,\bs p_1,s}\\{\bs{k},\zeta}}}\kappa_{\bs p \bs p_1}(\bs{k},\zeta)\left(\hat{b}_{\bs{k},\zeta}{+}\hat{b}_{\bs{k},\zeta}^\dagger\right)\hat{a}_{\bs p_1,s}^\dagger\hat{a}_{\bs p,s}}_{\rm electron-phonon\;interaction} \nonumber\\
    &{+}\underbrace{\sum_{\substack{\bs p \bs p_1\\\bs{q},\xi}}\left(\lambda_{\bs p \bs p_1}^-(\bs{q},\xi)\hat{c}_{\bs{q},\xi}+\lambda_{\bs p \bs p_1}^{+}(\bs{q},{\shortminus \xi})\hat{c}_{\bs{q},{\shortminus \xi}}^\dagger\right)\hat{a}_{\bs p_1,\xi}^\dagger\hat{a}_{\bs p,{\shortminus}\xi}}_{\rm electron-photon\;interaction}
    .
    \label{eq:hamiltonian}
\end{align}

Here $\bs q=\bs q_{\parallel}+q_z(l) \bs e_z$ is the photon 3D momentum with $\bs q_{\parallel}=q_x\bs e_x+q_y\bs e_y$ being the in-plane momentum. The transverse quantization level number is denoted by $l$. Hence, a momentum summation is over both the 2D-momentum and the transverse level, $\sum_{\bs q} = \sum_{\bs q_\parallel,l}$. The $\hat c_{\bs q,\xi}$ is the photon annihilation operator in a state with the momentum $\bs q$ and polarization $\xi$ ($\xi=\pm1$ for right/left circular polarization). Thus, the frequency is given as
\begin{align}
    \omega_{\bs q}=\frac{c}{\sqrt\varepsilon}\sqrt{q_z^2(l)+\bs q_\parallel^2},
\end{align}
where $\varepsilon$ is the dielectric constant of the medium inside the microcavity and $c$ is the vacuum speed of light. The condensate photon frequency is denoted by $\omega_c\equiv\omega_{\bs 0}=cq_z(0)/\sqrt\varepsilon$.

The same decomposition is adopted for the phonon momentum $\bs k=\bs k_{\parallel}+k_z \bs e_z$ with $\bs k_{\parallel}=k_x\bs e_x+k_y\bs e_y$. The operator $\hat{b}_{\bs k,\zeta}$ annihilates a phonon with ${\bs k}$ momentum in a mode $\zeta$ (acoustic and optical modes). Dispersions are given by $E_{\rm acoustic}(k)=ku$ and $E_{\rm optical}={\rm const}$, where $u$ is the speed of sound and $k=\sqrt{{\bs k}_\parallel^2+k_z^2}$.

For electrons, we use the standard expression \cite{LL3} for quantized levels in magnetic field:
\begin{align}
E_{\bs p}^s=\hbar\omega_H\left(n+\frac12\right)+\hbar\omega_z\left(\mathfrak{L}+\frac12\right)+\frac{\hbar\omega_s}2 s,
\end{align}
with $\omega_H = eH/m^*_ec$ being the cyclotron frequency, $\omega_z$ standing for the electron trap frequency in direction $Oz$, $s=\pm1$ denotes the sign of the electron spin projection. The Zeeman frequency $\omega_s = eH/m_ec\ll\omega_H$.

We use a collective index for the electron state: $\bs p_n\equiv \{p_x,n,\mathfrak{L}\}$: $p_x$ is the momentum projection on in-plane direction $Ox$, $n$ stands for the Landau level number, $\mathfrak{L}$ is the transverse quantization level number. Thus, $\hat{a}_{\bs p,s}$ annihilates a photon from trapping potential level $\mathfrak{L}$, Landau level $n$ with spin projection $s$. Note, that the electron spectrum does not depend on the $p_x$, thus we will further eliminate it.

Hereinafter we consider electrons from a single transverse quantization level, thus $\mathfrak{L}$ is set to zero and further omitted: $E_{n}^s\equiv E_{n,\mathfrak{L}=0}^s$.

As already mentioned, in subsequent sections we consider magnetic field with spatial gradient in order to facilitate energy absorption from the thermal source to the pumping modes. As estimated in Appendix \ref{app:gradient}, the gradient can be chosen weak enough not to affect the Hamiltonian, we just treat $\omega_H$ above as the average magnetic field across the area of the device.

The electron-photon and electron-phonon interaction constants are given by $\lambda_{\bs p \bs p_1}^\pm(\bs{q},\xi)$ and $\kappa_{\bs p \bs p_1}(\bs{k},\zeta)$ respectively.  For further calculations we will not use the whole set of indices in $\bs p$ since we consider electrons from $\mathfrak{L}=0$ and omit $p_x$:
\begin{align*}
    \lambda_{\bs p \bs p_1}^\pm(\bs{q},\xi)\equiv\lambda_{nm}^\pm(\bs q,\xi)\delta_{p_{x},p_{1x}}\\
    \kappa_{\bs p \bs p_1}(\bs{k},\zeta)\equiv\kappa_{n m}(\bs k, \zeta)\delta_{p_{x},p_{1x}}.
\end{align*}
The definitions of both couplings are given in Appendix \ref{app:couplings}. 

\section{Kinetics}\label{sec:kinetics}

\begin{figure*}[t]
\begin{center}
\includegraphics[]{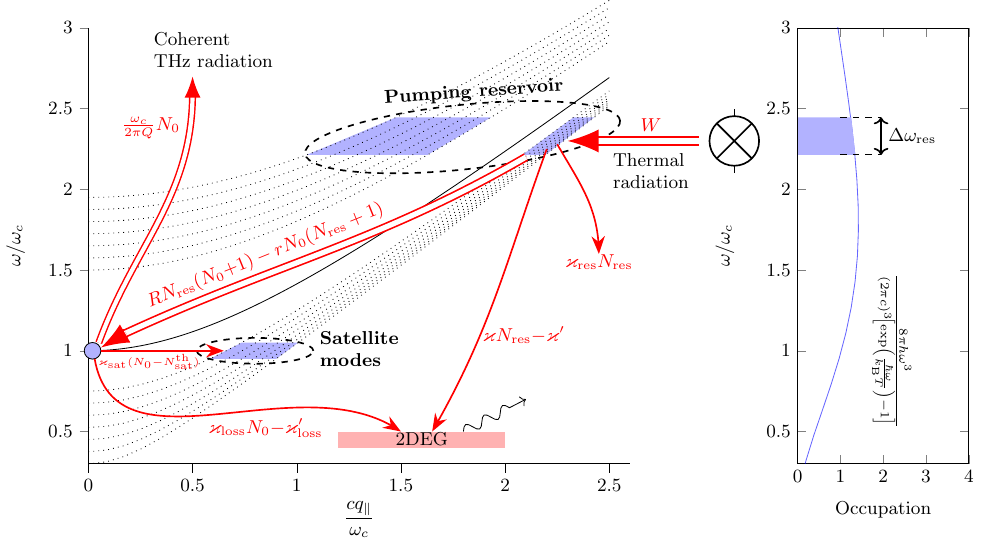}
\end{center}
\caption{A schematic of all the relevant processes in the photonic subsystem of interest (blue shaded regions). (a) Dotted lines on the $\omega(\bs q_\parallel)$ plane depict modes of the same transverse momentum(quantized). Red arrows depict particle exchange processes, with their orientation defining the positive sign convention. Double arrows highlight the dominant processes in the condensate state. The black wavy line depicts a phonon, which is emitted when 2DEG absorbs a photon. (b) Thermal radiation intensity frequency profile with the pumping region being shaded.}
\label{fig:kinetic_scheme}
\end{figure*}

In this section we investigate the kinetics of the photonic subsystem of interest by separately addressing all the relevant modes and calculating the rates of all the relevant processes, which are summarized in Fig. \ref{fig:kinetic_scheme}. Unless stated otherwise, we use the word "mode"\ for a state of photon with fixed total momentum, both it is in-plane and transverse components.

\subsection{Condensate mode}
In order to describe the kinetic processes, which feed/deplete condensate mode, we make use of the Keldysh technique ~\cite{LL10}. Rates of photonic exchange are given in terms of the retarded self-energy term $\Sigma^R(t-t')=\sigma^R\delta(t-t')$:

\begin{align}
    &\frac{\partial N_0}{\partial t}=\Im[\sigma^R] =\!\!\!\!\sum_{j\in \Big\{\substack{{\rm pump}\\{\rm loss}\\{\rm sat}}\Big\}}\left.\!\!\!\!\!\!\!d_t N_0\right|_{j}.
\end{align}

The index $j$ runs over all the possible photon scattering channels. For each of them we may express the corresponding term in the Fermi's golden rule form:
\begin{align}
    &\ \ \ d_t N_0\big|_{j}=\frac{2\pi}{\hbar}\sum_{\bs p,\bs p'}|A^{(j)}_{\bs p,\bs p'}|^2\delta\left(E^{(j)}_\text{initial}(\bs p)-E^{(j)}_\text{final}(\bs p')\right){\times}\nonumber\\
    &\!\!\!\!\left[N^{(j\, in)}_\text{init}(\bs p)N^{(j\, in)}_\text{final}(\bs p')(1{+}N_0){-}
    N^{(j\, out)}_\text{init}(\bs p)N^{(j\, out)}_\text{final}(\bs p')N_0\right].
    \label{eq:boltzmann}
\end{align}

Here we used the amplitudes $A^{(j)}_{\bs p,\bs p'}$, total energies $E^{(j)}_{\rm initial/final}$ and occupation number products $N^{(j\, in/out)}_{\rm init/final}$ of initial and final scattering states.

Now we separately address to all the relevant couplings of the condensate mode, with all the calculation details being deferred to the Appendix \ref{app:calculations}.

\subsubsection*{(i) Condensate replenishment from the pumping reservoir}

Particle exchange between the condensate and pumping reservoir is due to resonant scattering on electrons. The corresponding Feynman diagrams are given in Fig. \ref{fig:pump-diagram}: 
\begin{figure}[H]
    \begin{center}
            \includegraphics[width=\linewidth]{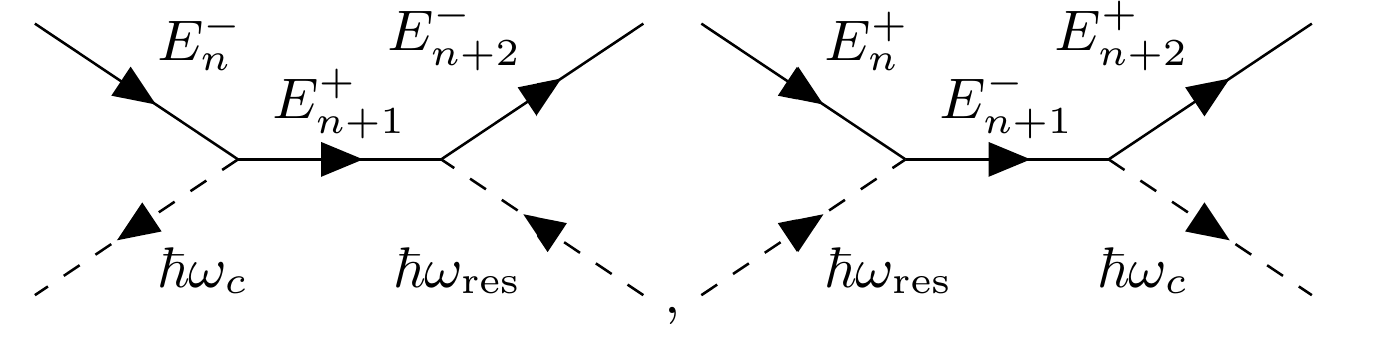}  
    \end{center} 
    \caption{Particle exchange diagrams for photons from the condensate and pump modes. Solid lines depict electrons, dashed ones -- photons. Indices $\pm$ stand for electron spin.}
    \label{fig:pump-diagram}
\end{figure}
Note that due to the magnetic field orientation in Fig. \ref{fig:scheme}, only left-polarized photons are involved in scattering processes due to angular momentum conservation (see Appendix \ref{subapp:electron-photon}).  Thus, hereafter we set $\xi=-1$ when considering electron-photon coupling $\lambda^\pm_{nm}({\bs q},\xi=-1)\equiv\lambda^\pm_{nm}(\bs q)$ and omit it.

The flux difference is:
\begin{equation}
    d_t N_0\Big|_{\rm 
    pump}\!\!\!= RN_{\rm res}(1+N_0)-r(1+N_{\rm res})N_0,
\end{equation}
In this equation and further we treat all the pumping modes collectively as a single reservoir with $N_{\rm res}$ being its occupation. By doing so, we assume that the pumping range broadening due to the magnetic field gradient is narrow enough for all the kinetic rates and the black body photon occupation not to change significantly over this interval.

The rates introduced above are given as follows:

\begin{align}
    &R{=}\frac{2\pi}{\hbar}\!\!\!\sum_{\substack{\bs p_n,{\bs q}_{\rm res}\\{s=\pm 1}}}\!\!\left|\frac{\lambda_{n+1\shortminus s,n+1}^-(\bs q_{\rm res})\lambda^{+}_{n+1,n+1+s}(\bs 0)}{E^{s}_{n+1+s}+s\hbar\omega_c-E_{n+1}^{-s}}\right|^2\nonumber\\
    &\times N^{\rm el}_{n,s}\left(1{-}N^{\rm el}_{n+2,s}\right)\delta(\hbar\omega_{\rm res}{+}E_n^s{-}\hbar\omega_c{-}E_{n+2}^s),
    \label{eq:nu-pump}
\end{align}
\begin{align}
    &r{=}\frac{2\pi}{\hbar}\!\!\sum_{\substack{\bs p_n, \bs q_{\rm res}\\{s=\pm 1}}}\!\left|\frac{\lambda^{-}_{n+1+s,n+1}(\bs 0)\lambda_{n+1,n+1\shortminus s}^+(\bs q_{\rm res})}{E^{s}_{n+1+s}+s\hbar\omega_c-E_{n+1}^{-s}}\right|^2\nonumber\\
    &\times N^{\rm el}_{n+2,s}\left(1{-}N^{\rm el}_{n,s}\right)\delta(\hbar\omega_{\rm res}{+}E_n^s{-}\hbar\omega_c{-}E_{n+2}^s).
    \label{eq:nu-reverse}
\end{align}
The 3D momentum of the pumping photon is denoted by $\bs q_{\rm res}$, electron level occupation is introduced as
\begin{equation*}
    N_{n,s}^{\rm el}=\left[\exp\left(\frac{E^{s}_n-\mu_{\rm el}}T\right)+1\right]^{-1}.
\end{equation*} 
with $T$ and $\mu_{\rm el}$ being the lattice temperature and electron chemical potential.

\subsubsection*{(ii) Phonon-assisted photon absorption by Landau-quantized electrons}

\begin{figure}[H]
\begin{center}
    \includegraphics[width=\linewidth]{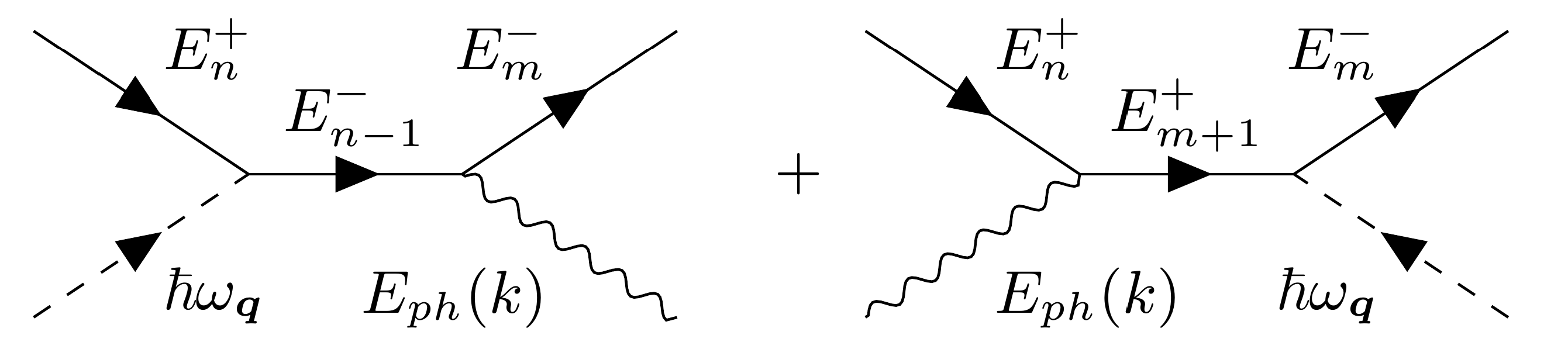}
\end{center}
\caption{Diagrams for photon absorption by the 2DEG with emission of phonon. Waved line depicts the phonon.}
    \label{fig:leak-diagram}
\end{figure}

The diagrams in Fig. \ref{fig:leak-diagram} describe photon absorption by the 2DEG with phonon emission.

The rates of spontaneous and stimulated processes for a photonic mode of 2D momentum $\bs q$ are given by the following expressions for $\varkappa_{\rm loss}'(\bs q)$ and $\varkappa_{\rm loss}(\bs q)$ respectively: 
\begin{widetext}
\begin{align}\label{eq:spontaneous_loss}
    \varkappa_{\rm loss}'(\bs q){=}\frac{2\pi}{\hbar}\!\!\sum_{\substack{p_x, \bs k\\{n,m,\zeta}}}\!
    \left|\!\frac{\lambda^{-}_{n,n\shortminus 1}(\bs q)\kappa_{n\shortminus1,m}(\bs k,\zeta)
    }{E^{+}_{n}{+}\hbar\omega_{\bs q}{-}E_{n\shortminus1}^{-}}{+}\frac{\kappa_{n,m+1}(\bs k,\zeta)\lambda^{-}_{m+1,m}(\bs q)
    }{E^{-}_{m}{-}\hbar\omega_{\bs q}{-}E_{m+1}^{+}}\!\right|^2\!\!\frac{N^{\rm el}_{m,\shortminus1}\left({1{-}N^{\rm el}_{n,+1}}\right)}{e^{E_{\rm ph}(\bs k,\zeta)/T}-1}
    \delta(\hbar\omega_{\bs q}{+}E_{n}^+{-}E_{m}^-{-}E_{\rm ph}({\bs k},\zeta)).
\end{align}
\begin{align}\label{eq:stimulated_loss}
    \varkappa_{\rm loss}(\bs q)&{=}\frac{2\pi}{\hbar}\!\!\sum_{\substack{p_x, \bs k\\{n,m,\zeta}}}
    \!\left|\!\frac{\lambda^{-}_{n,n\shortminus 1}(\bs q)\kappa_{n\shortminus1,m}(\bs k,\zeta)
    }{E^{+}_{n}{+}\hbar\omega_{\bs q}{-}E_{n\shortminus1}^{-}}{+}\frac{\kappa_{n,m+1}(\bs k,\zeta)\lambda^{-}_{m+1,m}(\bs q)
    }{E^{-}_{m}{-}\hbar\omega_{\bs q}{-}E_{m+1}^{+}}\!\right|^2\!\!\!\delta(\hbar\omega_{\bs q}{+}E_{n}^+{-}E_{m}^-{-}E_{\rm ph}({\bs k},\zeta))\nonumber\\
    &{\times}\left(\frac{N^{\rm el}_{n,+1}{-}N^{\rm el}_{m,\shortminus1}}{e^{E_{\rm ph}(\bs k,\zeta)/T}-1}+N^{\rm el}_{n,+1}\left(1{-}N^{\rm el}_{m,\shortminus1}\right)\right)
\end{align}
\end{widetext}

We further suppress the momentum argument for the condensate mode, thus $\varkappa_{\rm loss}=\varkappa_{\rm loss}(\bs q=0)$, $\varkappa'_{\rm loss}=\varkappa'_{\rm loss}(\bs q=0)$.

Evidently, the loss rate is maximized at the arguments of the energy-conserving delta-function, as qualitatively depicted in Fig. \ref{fig:cyclotron_idea}. The corresponding frequencies $\omega_{\rm c}$ are given by
\begin{align}\label{eq:acoustic_delta}
    \omega_c{=}\omega_{H}\Delta n_a+\omega_{s},\, \Delta n_a\geq0
\end{align}
for acoustic phonons and
\begin{align}\label{eq:optical_delta}
    \omega_c\pm\omega_{\rm opt}{=}\omega_{H}\Delta n_{\rm opt}+\omega_{s}
\end{align}
for optical ones. The quantities $\Delta n_{a/{\rm opt}}$ denote the number of Landau levels between the scattering states of an electron.

In Appendix \ref{subapp:loss rate} calculations are presented for the broadening of the peaks of the loss function.
\subsubsection*{(iii) Interaction with satellite modes}

A left-polarized photon from the condensate may elastically scatter off an electron to a satellite  mode of the same total energy. The process is given by diagrams in Fig. \ref{fig:mode-diagram}. The key feature of this process is its independence of occupation of the final modes:
\begin{figure}
\begin{center}
    \includegraphics[width=\linewidth]{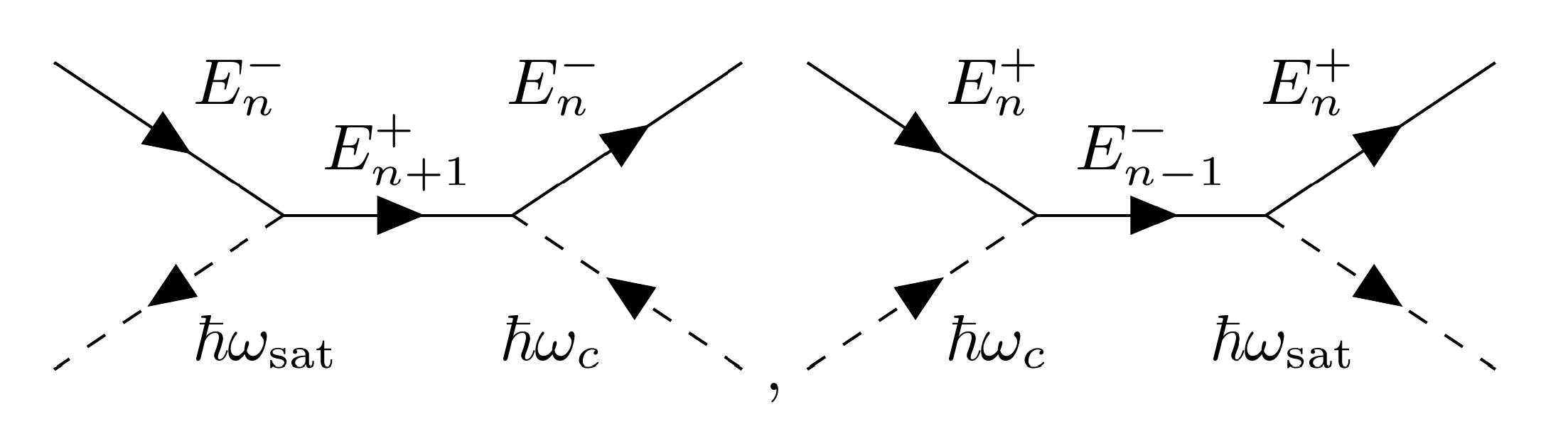}
\end{center}
    \caption{Diagrams for photon scattering from the condensate to a satellite mode with $\omega_{\rm sat}=\omega_c$ and nonzero $\bs q_{\rm sat}$.}
    \label{fig:mode-diagram}
\end{figure}

\begin{align}
    \varkappa_{\rm sat}&{=}\frac{2\pi}{\hbar}\!\!\!\sum_{\substack{\bs p_n\\\bs q_{\rm sat}\neq 0\\{s=\pm 1}}}\!\!\!\left|\!\frac{\lambda_{n,n\shortminus s}^-(\bs q_{\rm sat})\lambda^{+}_{n\shortminus s,n}(\bs 0)}{E^{s}_{n}{+}s\hbar\omega_c{-}E_{n\shortminus s}^{-s}}\right|^2\nonumber\\
    &\times N^{\rm el}_{n,s}\left(1{-}N^{\rm el}_{n,s}\right)\delta(\hbar\omega_{\rm sat}{-}\hbar\omega_c).
    \label{eq:nu-sat}
\end{align}
Here $\bs q_{\rm sat}$ denotes the 3D momentum of satellite mode photon.
\subsection{The pumping reservoir}
Besides being coupled to the condensate mode, the pumping reservoir is exposed to the black-body radiation, leakage through the mirrors and weak photon absorption/emission by the 2DEG (recall that the pumping region is chosen in the "absorption valley"\ with low $Q\sim 1$, see Fig. \ref{fig:cyclotron_idea}). The injection rate from the blackbody source is given by the following expression (here $\bs v$ is the photon velocity, $\bs S_{\rm res}$ is the oriented area of the sample, $\Delta \omega_{\rm res}$ is the width of the pumping region as shown in Fig. \ref{fig:cyclotron_idea}):
\begin{align}
    W &{=} \frac{\Delta N}{\Delta t} {=} \!\!\int\!\!\! \frac{d^3p}{(2\pi\hbar)^3}\frac{\bs v(\bs p)\cdot \bs S_{\rm res}}{\exp\left(\frac{\hbar\omega_{\bs p}}{T_\text{BB}}\right){-}1}\Theta\!\!\left(\frac{\Delta \omega_{\rm res}}{2}{-}|\omega_{\bs p}{-}\omega_{\rm res}|\right)\nonumber\\
    &=\varkappa_{\rm res}N_{\rm BE}(T_{\rm BB}),
\end{align}
whereas the losses through mirrors for the low Q region are given by 
\begin{align}
    \varkappa_{\rm res} = \frac{\Delta N}{N_{\rm BE}(T_{\rm BB})\Delta t} &{\approx} \frac{\pi \varepsilon cS_{\rm res} \omega_{\rm res}^2\Delta \omega_{\rm
    res}} {(2\pi c)^3}.
\end{align}
We introduced above $N_{\rm BE}(T_{\rm BB})=(\exp(\hbar\omega_{\rm res}/T_{\rm BB})-1)^{-1}$, representing the thermal occupation of the pumping reservoir that would result from the blackbody acting as the only source. 

The photon absorption/emission rates $\varkappa/\varkappa'$ are given by the same diagrams and expressions as $\varkappa_{\rm loss}/\varkappa'_{\rm loss}$ for the condensate mode (see Fig. \ref{fig:leak-diagram}).

Altogether: 
\begin{align}
    d_t N_{\rm res}&{=} W+\varkappa'{-}N_{\rm res}\big(\varkappa{+}\varkappa_{\rm res}\big)\nonumber\\
    &{+}R(1+N_0)N_{\rm res}-rN_0(1+N_{\rm res}).
     \label{eq:kinetic_pumping}
\end{align}
Here $\varkappa \equiv\varkappa_{\rm loss}(\bs q_{\rm res})$ and $\varkappa' \equiv\varkappa'_{\rm loss}(\bs q_{\rm res})$ (see eqs \eqref{eq:spontaneous_loss}-\eqref{eq:stimulated_loss}).


\subsection{Satellite modes}
As mentioned above, when describing condensate coupling to the satellite modes, the coupling rate does not depend on occupations of the latter. That is why we do not consider a kinetic equation for them. In all the further calculations we assume they are thermally populated with occupation $N_{\rm sat}^{\rm th}=N_{\rm BE}(\omega_c)$.
\subsection{Stationary state}
Altogether, the resonator photonic mode occupations are governed by the following ODE system: 
\begin{equation}\label{eq:full_kinetic}
\begin{cases}
 d_tN_0 \;\;
 {=}W_0+RN_{\rm res}{+}N_0\left(N_{\rm res}(R{-}r){-}r{-}\varkappa_0-\varkappa_{\rm mirror}\right),\\
d_tN_{\rm res}{=}W_{\rm res}{-}N_{\rm res}(R{+}\varkappa{+}\varkappa_{\rm res}){-}N_0\left(N_{\rm res}(R{-}r){-}r\right),
 \end{cases}
\end{equation}
where we introduced $W_0 = \varkappa_{{\rm loss}}'+\varkappa_{{\rm sat}}N_{\rm sat}^{\rm \rm th}$, $W_{\rm res} = W+\varkappa'\simeq W$, $\varkappa_0=\varkappa_{\rm loss}+\varkappa_{\rm sat}$ and $\varkappa_{\rm mirror}=\omega_c/2\pi Q$.

Stationary occupations of both the condensate and the reservoir are nonzero because of the spontaneous processes and finite due to finite influx $W$. Namely, for small $W=\varkappa_{\rm res}N_{\rm BE}(T_{\rm BB})$ and $W_0$, the occupations of both the condensate mode and the reservoir follow the thermal one:
\begin{equation}
    \begin{split}
        N_{\rm res}&{=}\frac{\varkappa_{\rm res}(\varkappa_0{+}\varkappa_{\rm mirror}{+}r)N_{\rm BE}(T_{\rm BB}){+}W_0r}{(\varkappa_0+\varkappa_{\rm mirror})R{+}(\varkappa{+}\varkappa_{\rm res})(\varkappa_0{+}\varkappa_{\rm mirror}{+}r)},\\
        N_0&{=}\frac{\varkappa_{\rm res}RN_{\rm BE}(T_{BB}){+}W_0(R{+}\varkappa{+}\varkappa_{\rm res})}{(\varkappa_0{+}\varkappa_{\rm mirror})R{+}(\varkappa{+}\varkappa_{\rm res})(\varkappa_0{+}\varkappa_{\rm mirror}{+}r)}.
    \end{split}
\end{equation}
In the opposite limit (for $W\to \infty$), the reservoir occupation is saturated at $N_{\rm max}$
\begin{equation}
\begin{split}
    N_{\rm res}&=N_{\rm max}\equiv\frac{r+\varkappa_0+\varkappa_{\rm mirror}}{R-r},\\
    N_0&=\frac{\varkappa_{\rm res}N_{\rm BE}(T_{\rm BB})-(\varkappa+\varkappa_{\rm res})N_{\rm max}}{\varkappa_0+\varkappa_{\rm mirror}}.
\end{split}\label{eq:infinity_limit}
\end{equation}

\begin{figure}[htp]
\begin{center}
\includegraphics[width = 0.9\linewidth]{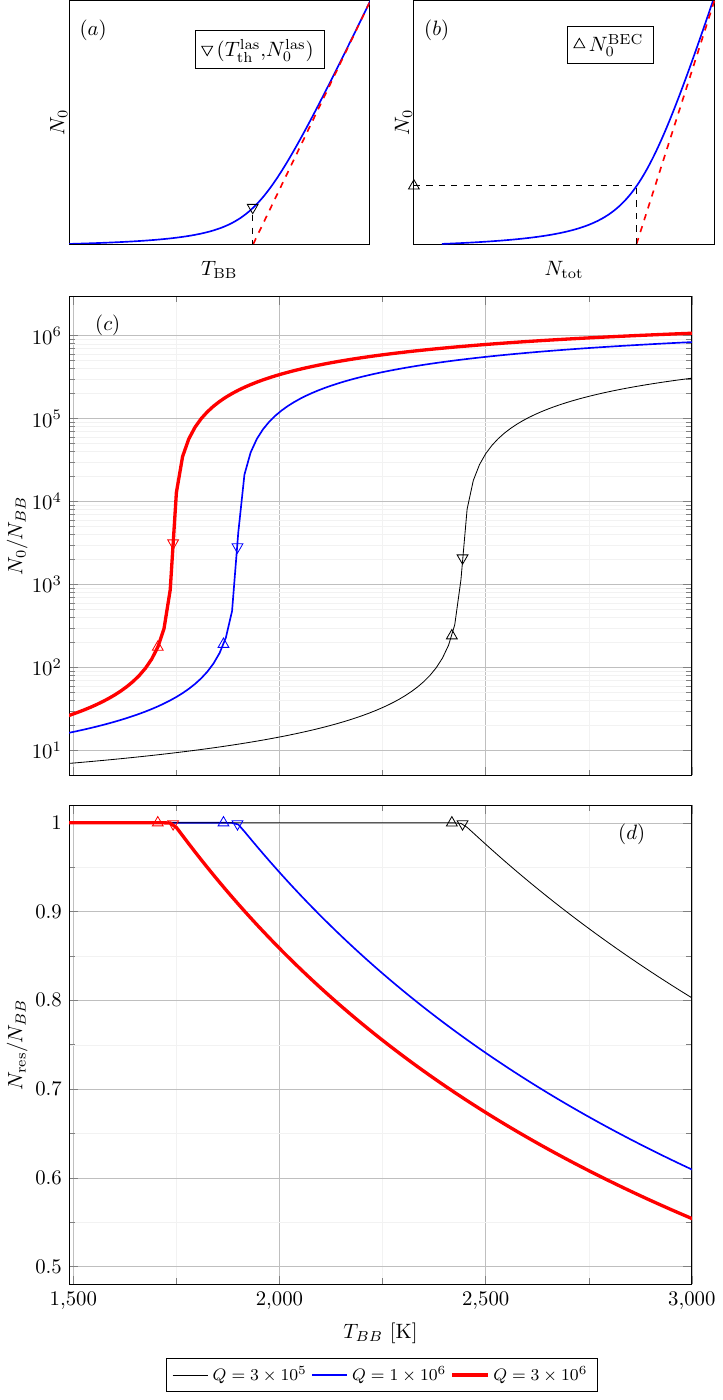}
\end{center}
\caption{Schematic representation of the procedures of finding (a) the critical lasing temperature; (b) the critical occupation for BEC formation. (c-d) The normalized condensate/reservoir occupation dependence on the blackbody temperature for fixed magnetic field $H=2.5$ T and various $Q$-factors. BEC transition points are indicated by upright triangles (\scalebox{0.9}[0.9]{$\triangle$}), while laser thresholds are marked by inverted triangles ($\triangledown$). For illustrative purposes, we use typical parameter values: $\varkappa_{\rm res}{=}2.1{\cdot}10^7\ \mu{\rm s}^{-1}$, $\varkappa=0.2\ \mu{\rm s}^{-1}$, $\varkappa_{\rm loss}{=}1.2\ \mu{\rm s}^{-1}$, $\varkappa_{\rm sat}{=}7.1$ $\mu{\rm s}^{-1}$, $R{=}4.4\ \mu{\rm s}^{-1}$, $r{=}1.8\ \mu{\rm s}^{-1}$, $\varkappa_{\rm sat}=7.1\ \mu{\rm s}^{-1}$, $\omega_{\rm c}=6.6$ meV and $\omega_{\rm res}{=}32.9$ meV which are estimated for a sample with an area $S\approx 300 $ $\mu$m at temperature $T=340$ K.}
\label{fig:occupations_vs_T}
\end{figure}

The exact solution of the system \eqref{eq:full_kinetic} for stationary states interpolates between the two regimes as demonstrated in Fig. \ref{fig:occupations_vs_T} (c)-(d).

When identifying the transition point between the two regimes we follow \cite{Semicond_BEC_lasing} to introduce the lasing and the Bose condensation thresholds $T_{\rm th}^{\rm las/BEC}$ (and the corresponding occupations $N_{0}^{\rm las/BEC}$). The lasing threshold is defined by extrapolating the linear dependence of the condensate occupation on the blackbody temperature $T_{\rm BB}$ as shown on panel (a) in Fig. \ref{fig:occupations_vs_T}. From \eqref{eq:infinity_limit} one may infer
\begin{align}\label{eq:lasing_th}
    N_{BE}(T^{\rm las}_{\rm th})=\left(\frac{\varkappa}{\varkappa_{\rm res}}+1\right)\frac{r+\varkappa_0+\varkappa_{\rm mirror}}{R-r}.
\end{align}

The BEC threshold is defined by extrapolating the linear dependence of the condensate occupation as a function of the total occupation (of the isolated photonic subsystem: condensate mode, satellite modes and the pumping reservoir) as demonstrated in Fig. \ref{fig:occupations_vs_T} (b). One may infer the value $N_0^{\rm BEC}$ from the graph and find the corresponding transition temperature, which is defined by
\begin{align}
    N_{\rm 0}(T_{\rm th}^{\rm BEC})=N_0^{\rm BEC}.
\end{align}

\section{Estimating the characteristics of the THz emitter}\label{sec:analysis}
Having investigated the physical principles of operation of the THz source, we address here to several details of manufacturing a physical device and assess its performance in terms of the output power and efficiency.
\subsection{Device engineering}
\subsubsection{Electron mobility}\label{subsec:mobility}

As stated in Section \ref{sec:idea}, the first step is engineering a cavity with an isolated high Q condensate mode and a bunch of pumping modes in the minima of the phonon-assisted photon absorption rate. The primary condition for this is the very existence of these minima, which requires ensuring high enough electron mean-free path in the 2DEG layer. To estimate the minimal required electron mobility for that, we use the kinetic equation for the condensate mode \eqref{eq:full_kinetic} and consider the condensate mode located at the center between absorption peaks of finite broadening due to limited electron mobility:
\begin{align}
    \Gamma=\frac{e\hbar}{\mu_em^*_e}.
\end{align}

As a candidate material for creating the heterostructure inside the microcavity as a support for 2DEG, we choose In$_{1-x}$Ga$_{x}$Sb with $x$ being modulated to create a trapping potential for electrons. This choice is motivated by one the highest electron mobilities $\mu_{\rm InSb}= 7.8\cdot 10^4$ cm${}^2$/V$\cdot$s \cite{Jubair24} and low effective mass $m^*_e/m_e = 0.015$ \cite{Levinshtein1996-gv}, which is beneficial for achieving higher cyclotron frequencies.

\begin{figure}[htp]
    \begin{center}
    \includegraphics[width = \linewidth]{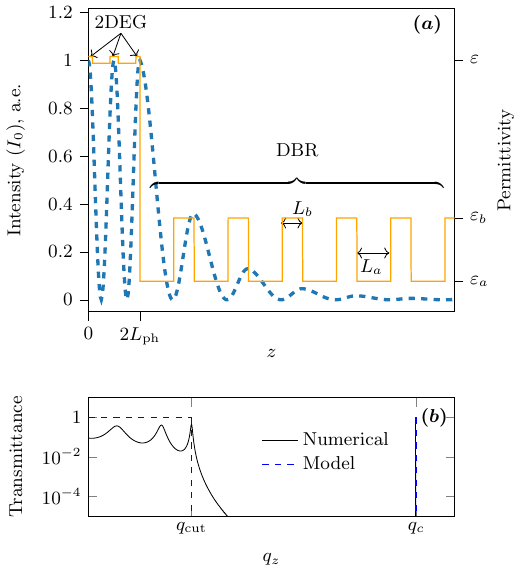}
    \end{center}
    \caption{(a) Resonator dielectric constant modulation in transverse direction (orange, right axis) and photon intensity profile (blue, dashed, left axis) for the condensate mode with the highest enhancement; (b) Resonator transmittance as a function for transverse photon momentum.}
    \label{fig:resonator_profile}
\end{figure}

We evaluate the critical broadening which increases the loss rate in the minimum enough to prevent condensate formation even in an ideal cavity with $Q\to \infty$. The magnetic field which corresponds to the critical broadening at room temperature $T=300$ K is estimated as $H_{\rm crit}\approx 2.3$ T (see Appendix \ref{app:broadening}). In further calculations we consider only $H>H_{\rm crit}$.

\subsubsection{Designing a cavity}

 To create the desired mode configuration, as described in Sec. \ref{sec:idea}, we propose using a heterostructure with high electron mobility inside a Bragg-reflector. This heterostructure hosts several layers of 2DEG (using multiple layers is helpful for increasing overall electron density). A possible pattern of the dielectric constant modulation is depicted in Fig. \ref{fig:resonator_profile} (a) (half of the structure is shown with a symmetric counterpart implied to the left). In the same plot we show the typical light intensity profile into the properly designed cavity.

The figure shows slight modulation of dielectric constant with respect to $\varepsilon$ inside the cavity, which is inevitable when creating a trap for the 2DEG. Though, we assume it to be small and ignore in our further calculations.

The key features of the resonator, which define the mode configuration are the lengths $L_{\rm ph}=\pi c/(\sqrt{\varepsilon}\omega_c)$, $L_{a/b}=\pi c/(\omega_c\sqrt{\varepsilon_{a/b}})$ where $\omega_c$ is the desired frequency of the condensate mode with high $Q$-factor. The latter depends on the length of the entire periodic structure. When designing a cavity, one has $\varepsilon$, $\varepsilon_{\rm a,b}$ and all the lengths under control. 

The typical form of the transmittance spectrum of the described structures presented on panel (b). Note the existence of a cutoff momentum $q_{\rm cut}$, which appears to significantly affect the results of calculation of kinetic rates. Its value is straightforwardly expressed from the equation (see \cite{Kavokin2011-vf}):
\begin{equation}
    \left|\frac{T_{11}+T_{22}}{2}\right|\le1,
\end{equation}
where $T_{11/22}$ are the diagonal elements of transfer matrix for 1D wave propagation.
The solution is as follows:
\begin{equation}
    \frac{q_{\rm cut}}{q_c}=\frac{2}{\pi}\arcsin\left(\frac{2\sqrt[\mbox{4}]{\varepsilon_a\varepsilon_b}}{\sqrt{\varepsilon_a}+\sqrt{\varepsilon_b}}\right),
\end{equation}
where $q_{{\rm cut},c} = \omega_{{\rm cut},c}\sqrt{\varepsilon}/c$.

For the sake of simplicity, the transmittance profile on panel (b) is approximated by a step function + a delta peak for further calculations. We use the following parameters: $\varepsilon = 15.2$ (InSb), $\varepsilon_a=2.1$, $\varepsilon_b=5.9$ (TiO$_2$/SiO$_2$ Bragg reflector). We consider cavities with $L_{\rm ph}\in [4,40] \, \mu$m, which corresponds to $\omega_c\in [1, 10]$ THz.

\subsection{Operating regimes}

\begin{figure}[htp]
    \begin{center}
    \includegraphics[width = \linewidth]{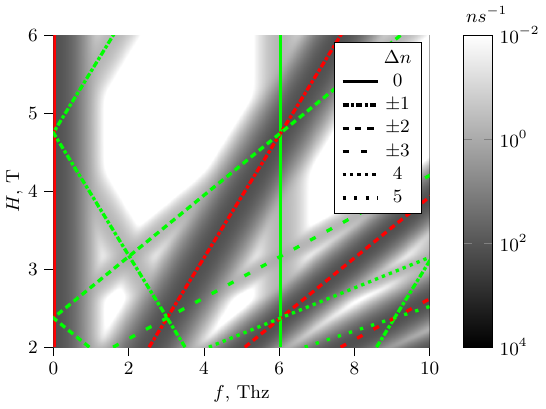}
    \end{center}
    \caption{The loss rate due to absorption of photons by the 2DEG. The light regions are the ones with small losses. The red lines correspond to the peaks of loss processes assisted by acoustic phonons, the green ones are for processes, which involve optical phonons. The type of the line used is determined by $\Delta n$, the difference of Landau level number of the scattering states of the electron.}
    \label{fig:leak-map}
\end{figure}

In order to macroscopically populate the condensate mode, one should adjust the control parameters of the system to ensure low level of photon absorption by the 2DEG. The Fig. \ref{fig:leak-map} shows the corresponding rate as a function of condensate frequency $\omega_c$ (which is defined by the cavity geometry) and the magnetic field, which may be varied in experiment. Optical phonon frequency ($\approx 6$ THz) is the one of InSb.

The lines of peak losses, which are given by \eqref{eq:acoustic_delta} and \eqref{eq:optical_delta} isolate several valleys of low absorption where condensate formation is possible provided high enough pumping rate. Note that even the lowest peaks are by orders of magnitude higher than any realistic pumping rate.

\begin{figure}[htp]
    \begin{center}
    \includegraphics[width=0.5\textwidth]{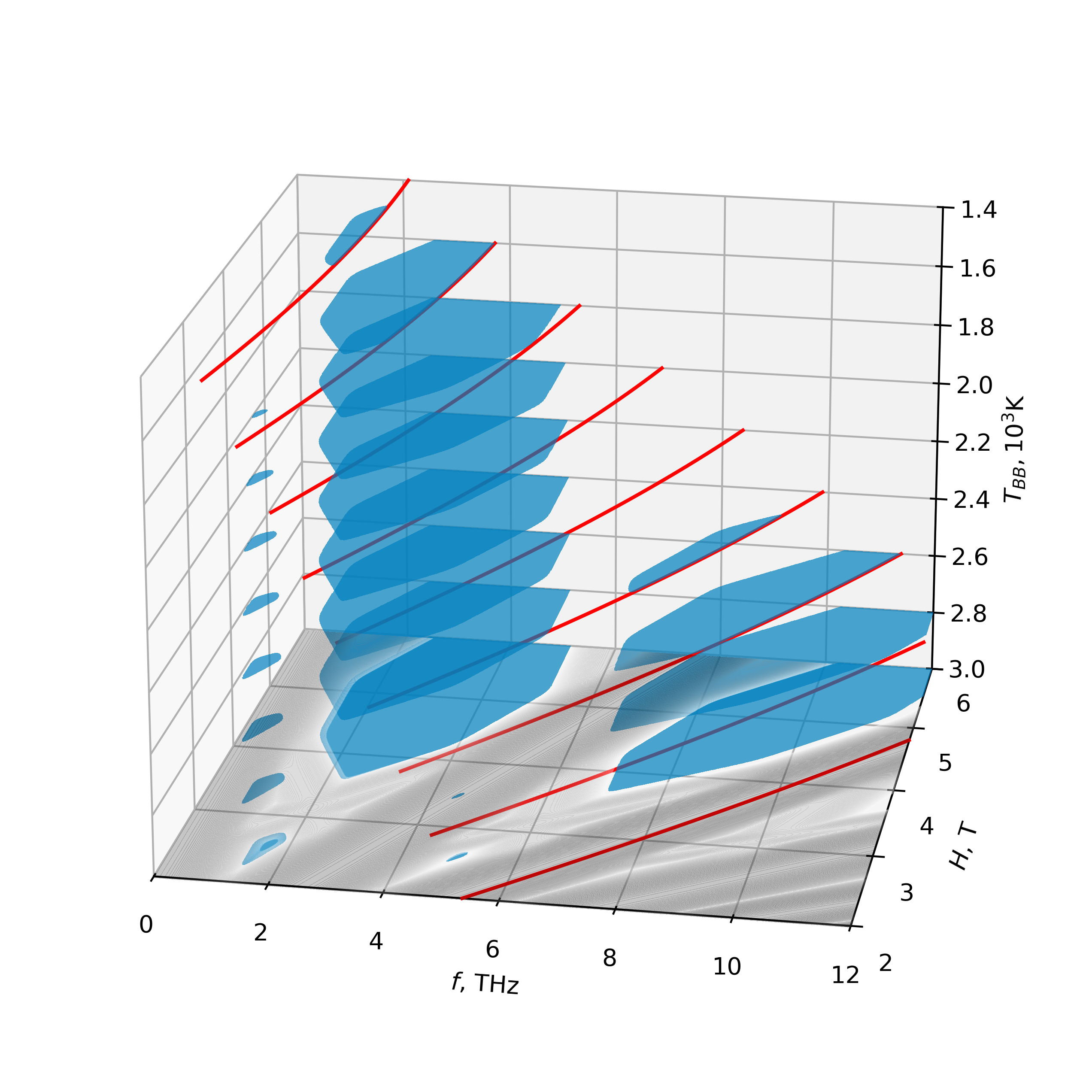}
    \end{center}
    \caption{A phase diagram of the system under consideration. In the regions with blue shading condensate is formed. The red lines show the critical pumping threshold given by the solutions of equation \eqref{eq:critical_pump}.}
    \label{fig:3d-temperature-map}
\end{figure}

The Fig. \ref{fig:3d-temperature-map} shows the regions of condensate existence for various temperatures of the thermal emitter in case of $Q\to \infty$ on the condensate frequency. Note that these regions (in blue) are bounded by both the absorption peaks and the line of threshold pumping level (in red), which is given as a result of solving the equation (see \eqref{eq:full_kinetic})

\begin{align}\label{eq:critical_pump}
    N_{\rm res}\!\big(R(\omega_c,\!H){-}r(\omega_c,\!H)\big){=}\varkappa_{0}(\omega_c,\!H){+}r(\omega_c,\!H){+}\varkappa_{\rm mirror}
\end{align}

One may clearly see how the area of the condensate phase shrinks with decreasing temperature (note the reversed axis). The critical temperature is $T_{\rm BB}^{\rm crit}=1300$ K (recall that the lattice temperature is $~300$ K).

\begin{figure}[htp]
\begin{center}
    \includegraphics[]{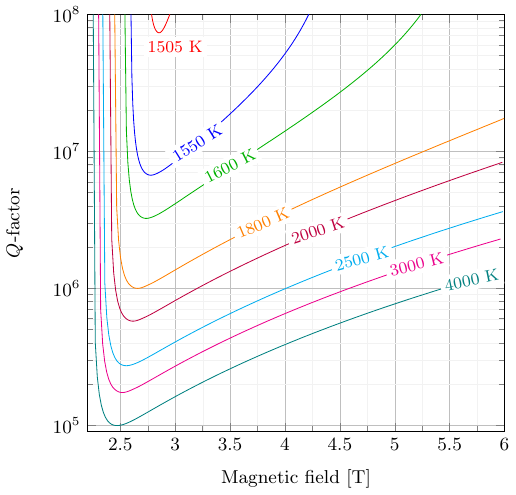}
\end{center}
\caption{The lasing threshold lines for various pumping temperatures. For each specific temperature the lasing state is above the corresponding line.}
\label{fig:criticalQH}
\end{figure}

No real mirror is perfect, which limits the usage of all the regions in Fig. \ref{fig:3d-temperature-map}. Fig. \ref{fig:criticalQH} shows the lasing threshold, where $N_0=N_{\rm BE}(T_{\rm th}^{\rm las})$ (see \eqref{eq:lasing_th}), helping to assess the feasibility of a real light-emitting device for various combinations of control parameters. Notably, neither low nor high magnetic fields are optimal. With decreasing magnetic field the relative width of the absorption minima (see Fig. \ref{fig:cyclotron_idea}) decreases, thus the losses increase. In contrast, higher magnetic fields shift the pumping region to higher frequencies, where the blackbody spectral density is exponentially suppressed. Therefore, an intermediate optimal magnetic field exists, enabling device operation with a low $Q$ -factor. 

One may see how the minimal required blackbody temperature decreases with increasing quality of the mirrors along with a wider range of magnetic fields. For this diagram we considered the largest region in Fig. \ref{fig:3d-temperature-map} (since it has the lowest possible operating temperature).

\subsection{Device performance}\label{subsec:device perform}

Now we study the device performance in terms of emission power and efficiency. Since the useful THz radiation is the one that escapes the microcavity through the mirror, the stationary emission intensity $I_c = \hbar\omega_c\cdot d_tn_0\big|_{\rm mirror}$ is as follows:
\begin{equation}
    I_c = \frac{W\hbar\omega_c\varkappa_{\rm mirror}}{\varkappa_{0}+\varkappa_{\rm mirror}}\left(1-\frac{N_{\rm max}}{N_{\rm BE}(T_{\rm BB})}\right).
\end{equation}

To estimate the input power to the device, we consider the energy flux from the black body in the same fashion as we estimated the particle flux:
\begin{align}
    I_{\rm in} \simeq \hbar\omega_{\rm res}W\left(1-\frac{N_{\rm max}}{N_{\rm BE}(T_{\rm BB})}\right).
    \label{eq:device_power}
\end{align}

Hence, the efficiency is given as:
\begin{equation}
    \eta = \frac{I_c}{I_{\rm in}}\simeq \frac{\omega_c}{\omega_{\rm res}}\frac{\varkappa_{\rm mirror}}{\varkappa_{0}+\varkappa_{\rm mirror}}.
    \label{eq:device_efficiency}
\end{equation}

\begin{figure}[htp]
        \includegraphics[scale = 0.54]{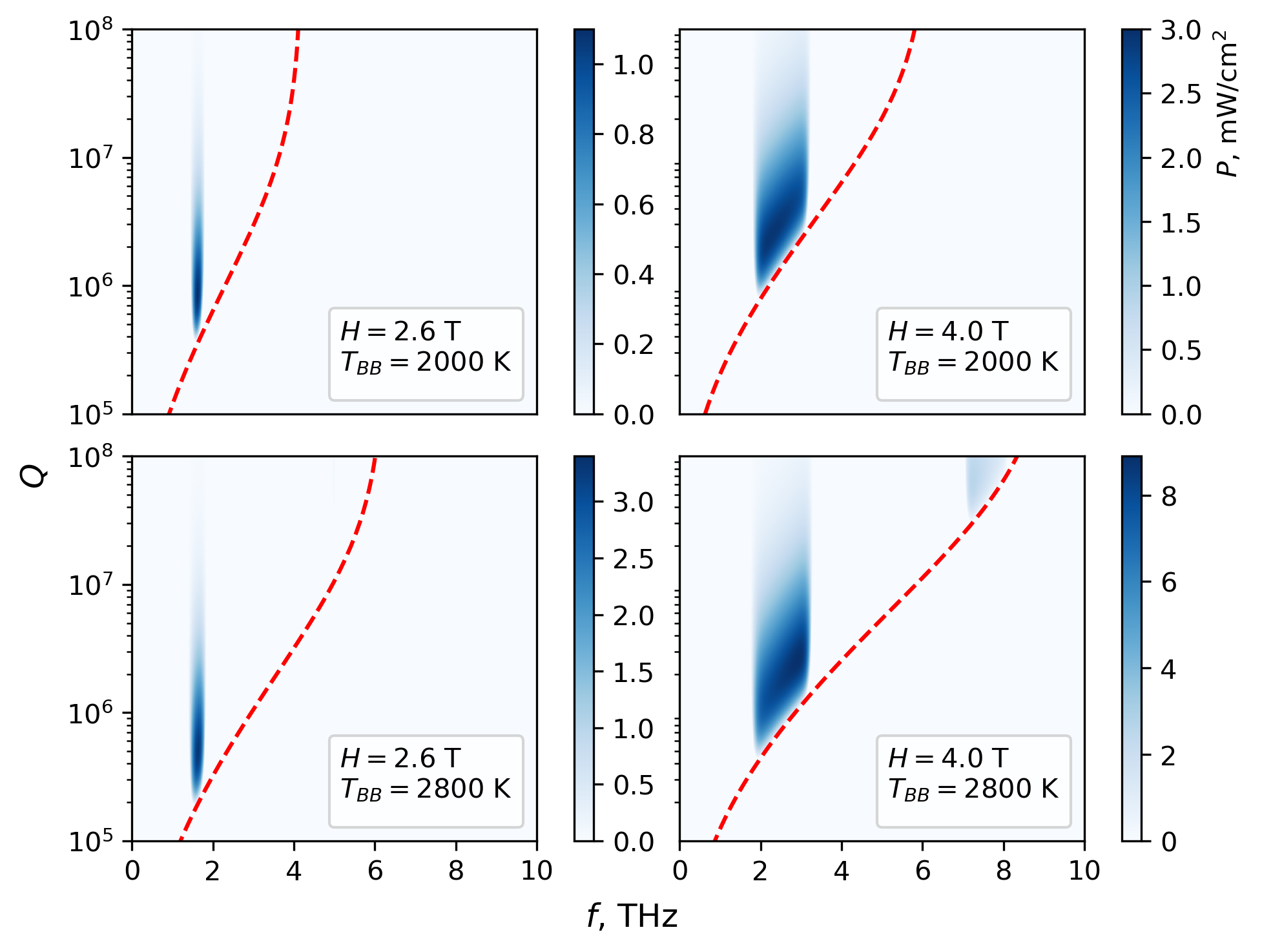}
        \caption{The power of the emitted light as a function of the emission frequency and cavity quality factor $Q$ at fixed magnetic field and temperature. The red dashed lines denote the threshold values  $Q_{\rm min}$, below which condensate is totally depleted.}
        \label{fig:device_power_Q_H}
    \end{figure}

In order to optimize the power and the efficiency of the THz emitter, one may vary the magnetic field as well as the cavity $Q$-factor. We calculate both the power and the efficiency using expressions \eqref{eq:device_power} and \eqref{eq:device_efficiency}. In Fig. \ref{fig:device_power_Q_H} the power output is shown as a function of the resonant frequency of the cavity $\omega_c$ and the corresponding $Q$-factor.

Clearly, for a low $Q$-factor, photon leakage out of the cavity is not compensated by pumping, thus no condensate will be present in the cavity at all. The threshold value is
\begin{equation}
    Q_{\rm min}=\frac{\omega_c}{2\pi(R-r)(N_{\rm BE}(T_{\rm BB})-N_{\rm max})}  
\end{equation}
and is shown by a red dashed line in the figure.

Operation region of the device is given by thin stripes in the figure \ref{fig:device_power_Q_H}. These are exactly the valleys between the peaks of condensate photon absorption by the 2DEG as depicted in Fig. \ref{fig:leak-map}. These stripes widen with the increase of magnetic field.

The maximal output power is achieved at some finite distance from the threshold $Q$-factor. Clearly, this is a result of an interplay of two effects: the higher $Q$-factor leads to higher condensate density $n_0$, albeit the lower one insures strong coupling to the environmental photonic modes.

Besides the absolute value of the output power, one is usually interested in the efficiency of the light source. As an upper bound, we calculate the efficiency, considering only the energy influx to the cavity in pumping frequency range as shown in Fig. \ref{fig:kinetic_scheme} (b).  It is exactly the expression \eqref{eq:device_efficiency}, which leads to an estimate of 25\%.

Clearly, the practical implementation of a THz-emitting device of the type described in this paper will face multiple challenges. Technical details may strongly affect the final outcome. We outline a few considerations:
\begin{enumerate}
    \item The threshold $Q$-factor is quite high compared to those achieved experimentally  ($\sim10^3$)\cite{Yu18}. However, as described in the same Ref. \cite{Yu18}, adding several more layers to the Bragg reflectors may increase $Q$ by orders of magnitude. In addition, one is not constrained by thermal pumping, which we use for demonstrative calculations. Non-equilibrium pumping sources may be used, such as photonic crystal structures that increase the flux on the desired pumping frequency. With increased pumping flux, the required $Q$-factor can be reduced.

    \item Another way to increase the thermal input to the cavity is to use THz (metallic) mirrors that reflect the radiation from undesired directions and redirect it into the cavity.

    \item In device design, it is desirable to minimize the energy consumed by the lamp to produce radiation at higher frequencies. This may be achieved by placing the lamp inside a cavity made of high-frequency (IR frequency, visible light) Bragg mirrors, which trap the waste radiation. Thus, the power consumption of the lamp decreases once it reaches equilibrium with the cavity.
\end{enumerate}

\section{Conclusions}\label{sec:conclusion}

In this paper we present a new mechanism for observing microcavity photonic BEC in the THz range. We studied a system of microcavity photons exposed to pumping by a thermal light source. The temperature difference between the cavity and the pumping source drives a photonic flux to the former. Due to interphoton interactions via Landau-quantized electrons in a magnetic field, the system settles down in a stationary state with macroscopic occupation of a single transverse mode of the cavity.

The demonstrated ability to populate a single transverse mode is necessary though insufficient condition for achieving full coherence. In contrast with cold-atom condensate, there is no direct spontaneous symmetry breaking because of negligible interaction between condensate photons. The condensate gain saturation mechanism is purely kinetic (limited influx from the pumping modes), which is closer to the condensation scenario in exciton-polariton condensates.

In this study we treated a set of low-lying modes of the condensate collectively. This definitely precludes the essential discussion on the thermalization thermalization among states with the same transverse momentum and statistics of the condensate. That is why the estimation of the coherence degree of the photonic condensate and investigation of its self-thermalization dynamics are beyond the scope of the current paper. A detailed study of this issue (along the lines of \cite{PhysRevLett.108.160403} for the photonic condensate in a dye-filled cavity) is a subject of future investigations.

Altogether, we expect this work to motivate experimental realizations of terahertz photon condensate. Room-temperature photonic BEC of the type described here is ostensibly useful for both fundamental investigations and applications as a sought-after coherent light source in the THz region.

\section{Acknowledgments}
We express our acknowledgment to A.P. Shkurinov for the the key proposition of using magnetic field to suppress photonic loss. We thank S.E. Kuratov and W.V. Pogosov for their interest in the long-term calculations for this paper and for useful discussions of the results. The work of N.A.A., I.L.K. and Yu.E.L. was supported by the Russian Science Foundation grant No. 23-42-10010, https://rscf.ru/en/project/23-42-10010/. The calculations of transmission spectra of Bragg reflectors were performed by I.L.K. within the framework of the topic FFUU-2024-0003.
\clearpage
\appendix
\section{Couplings}\label{app:couplings}
\subsection{Electron-phonon}\label{subapp:electron-phonon}
The electron-phonon coupling constant used in \eqref{eq:hamiltonian} in the main text is given as follows (see \cite{Snoke2007-we}):
\begin{align}
    \kappa_{mn}(\bs k)&=\sqrt{\frac{D_e^2 
    |k|}{2\rho u V}}V^{\parallel}_{mn}\left(\frac{k_{\parallel}^2l_H^2}{2\hbar^2}\right)V^{\perp}_{00}\left(\frac{k_z^2l_z^2}{2\hbar^2}\right).
    \label{eq:overlap}
\end{align}
The deformation potential is given by $D_e$, $\rho$ is the density of the semiconductor and $u$ is the speed of sound, $V$ is the volume of the system with periodic boundary conditions. We use $l_H$ and $l_z$ as the characteristic length of the electron potential $l_{H/z} = \sqrt{\hbar/m^*\omega_{H/z}}$.

The overlap integrals of Landau electron wavefunctions with the one of acoustic phonon are given as follows:
\begin{align*}
    V^{\perp}_{nm}(k_z)&\sim\int dz \psi^z_m(z)e^{ik_zz/\hbar}\psi_n^z(z),\\
    V^{\parallel}_{nm}(k_{\parallel})&\sim\int dy \psi^y_m(y,p_x')e^{ik_yy/\hbar}\psi_n^y(y,p_x)\Big|_{p_x'-p_x=k_x},
\end{align*}
with phonon wavefunction $\exp(i\bs k\bs r/\hbar)$ and electron wavefunctions ($H_m(x)$ stands for the $m^{\rm th}$ order Hermite polynomial):
\begin{align*}
    \psi^z_m(z)&=\frac{H_m(z/l_z)e^{-z^2/2l_z^2}}{\sqrt{2^mm!\sqrt{\pi}l_z}},\\
    \psi^y_m(y,p_x)&=\frac{H_m(\eta)e^{-\eta^2}}{\sqrt{2^mm!\sqrt{\pi}l_H}}.
\end{align*}

Here we use the shifted coordinate $\eta={y}/{l_H}-{p_xl_H}/{\hbar}$ as explained in \cite{LL3}.
As a result:
\begin{align}\label{eq:overlap_integral}
V^{\perp}_{mn}(x)=V^{\parallel}_{mn}(x)=\sqrt{\frac{n!}{m!}x^{m-n}}L_n^{(m-n)}(x)e^{-x/2},\ m\ge n
\end{align}
where $L_n^{(a)}(x)$ is generalized Laguerre's polynomial.

Evidently for phonons with momentum, which is significantly different from $\hbar/l_H$, the overlap integral $V_{mn}(x)$ tends to zero. That is the reason why absorption rate has periodic profile as shown in Fig.\ref{fig:cyclotron_idea}.

\subsection{Electron-photon}\label{subapp:electron-photon}
The electron-photon interaction is given by the standard Hamiltonian
\begin{equation*}
    \hat{H}_{e-ph}{=}\frac{e}{m^*_ec}\!\!\int \!\!d^3\bs{r}\hat\psi^\dagger(\bs r)\!\!\left[\!\hat{\bs A}(\bs r)\!\left(\hat{\bs p}{-}\frac{eH}{c}y\bs{e}_x\!\right)\!\right]\!\hat{\psi}_(\bs r).
\end{equation*}
The coupling constant with explicitly separated contributions for left and right polarized photons is as follows:

\begin{align}
    \lambda^{-}(\bs q,\xi)&{=}\hbar\omega_H{\psi}_{\rm ph}^j(l)\sqrt{\!\frac{e^2}{\varepsilon\hbar\omega_{\bs q}}\frac{2\pi l_H^2}{S}}\!\Bigg(\!\sqrt{{\rm max}(m,n)}\delta_{\mathfrak{L}\mathfrak{L}'}{\times}\nonumber\\
    &{\times}\left[\frac{\epsilon_{\bs q,\xi}^x{-}i\epsilon^y_{\bs q,\xi}}{\sqrt{2}}\delta_{m,n{+}1}{+}
    \frac{\epsilon_{\bs q,\xi}^x{+}i\epsilon^y_{\bs q,\xi}}{\sqrt{2}}\delta_{m,n{-}1}\right]\nonumber\\
    &{+}\!\frac{l_H}{l_z}\sqrt{{\rm max}(\mathfrak{L},\mathfrak{L}')}\delta_{mn}\frac{i\epsilon^z_{\bs q,\xi}}{\sqrt{2}}\left[\delta_{\mathfrak{L}',\mathfrak{L}{+}1}{-}\delta_{\mathfrak{L}',\mathfrak{L}{-}1}\right]\Bigg),
\end{align}
\begin{align}
    \lambda^{+}(\bs q,\xi)&{=}\hbar\omega_H{\psi}_{\rm ph}^j(l)\sqrt{\!\frac{e^2}{\varepsilon\hbar\omega_{\bs q}}\frac{2\pi l_H^2}{S}}\!\Bigg(\!\sqrt{{\rm max}(m,n)}\delta_{\mathfrak{L}\mathfrak{L}'}{\times}\nonumber\\
    &{\times}\left[
    \frac{\epsilon_{\bs q,\xi}^{x*}{-}i\epsilon^{y*}_{\bs q,\xi}}{\sqrt{2}}\delta_{m,n{+}1}{+}\frac{\epsilon_{\bs q,\xi}^{x*}{+}i\epsilon^{y*}_{\bs q,\xi}}{\sqrt{2}}\delta_{m,n{-}1}\right]\nonumber\\
    &{+}\!\frac{l_H}{l_z}\sqrt{{\rm max}(\mathfrak{L},\mathfrak{L}')}\delta_{mn}\frac{i\epsilon^{z*}_{\bs q,\xi}}{\sqrt{2}}\left[\delta_{\mathfrak{L}',\mathfrak{L}{+}1}{-}\delta_{\mathfrak{L}',\mathfrak{L}{-}1}\right]\Bigg).
\end{align}
Here $S/2\pi l_H^2$ is the degeneracy of each Landau level, $\bs{\epsilon}_{\bs q,\xi}$ stands for photon polarization, $\psi^j_{\rm ph}(l)$ is the amplitude of wave function of the photon with $l^{\rm th}$ mode in the $j^{\rm th}$ electron layer.
The only difference between the $\lambda^{\pm}$ is the complex conjugation of the polarization vector. Hereafter we neglect the third term of both expressions due to its smallness compared to the sum of Landau level. Thus, one may unify:
\begin{align}
    \lambda^\pm(q,\xi)&{=}\hbar\omega_H{\psi}_{\rm ph}^j(l)\sqrt{\!\frac{e^2}{\varepsilon\hbar\omega_{\bs q,l}}\frac{2\pi l_H^2}{S}{\rm max}(m,n)}{\times}\nonumber\\
    &{\times}\!\left[\!\!
    \left(\frac{\epsilon_{\bs q,\xi}^{x}{\pm}i\epsilon^{y}_{\bs q,\xi}}{\sqrt{2}}\right)^{\pm}\!\!\!\!\!\delta_{m,n{+}1}{+}\!\left(\frac{\epsilon_{\bs q,\xi}^{x}{\mp}i\epsilon^{y}_{\bs q,\xi}}{\sqrt{2}}\right)^{\pm}\!\!\!\!\!\delta_{m,n{-}1}\right].
\end{align}
Here the $\pm$ in the superscript stands for the presence/absence of complex conjugation.

For the sake of convenience, we pass to circular polarization. Foremost, arbitrary momentum $\bs q$ may be represented as a result of rotation of $\bs e^z$:
\begin{widetext}
\begin{equation}
    \bs q \doteq\begin{pmatrix}
  q\cos\phi\sin\theta\\
  q\sin\phi\sin\theta\\
  q\cos\theta
\end{pmatrix}=\begin{pmatrix}
  \sin\phi & \cos\phi & 0 \\
  -\cos\phi & \sin\phi & 0 \\
  0 & 0 & 1
    \end{pmatrix}\begin{pmatrix}
  1 & 0 & 0 \\
  0&\cos\theta & \sin\theta \\
  0& -\sin\theta & \cos\theta \\
\end{pmatrix}\begin{pmatrix}
  0\\
  0\\
  q
\end{pmatrix}.
\end{equation}
Thus, circular polarization vectors are given as follows ($\xi=\pm 1$ for right and left polarizations respectively):
\begin{equation}
    \bs \epsilon_{\bs q,\xi} =\begin{pmatrix}
  \epsilon^x_{\bs q,\xi}\\
  \epsilon^y_{\bs q,\xi}\\
  \epsilon^z_{\bs q,\xi}
\end{pmatrix}\doteq \begin{pmatrix}
   \sin\phi & \cos\phi & 0 \\
  -\cos\phi & \sin\phi & 0 \\
  0 & 0 & 1
    \end{pmatrix}
    \begin{pmatrix}
  1 & 0 & 0 \\
  0&\cos\theta & \sin\theta \\
  0& -\sin\theta & \cos\theta \\
\end{pmatrix}\begin{pmatrix}
  1/\sqrt2\\
  i\xi /\sqrt2\\
  0
\end{pmatrix}=\frac{1}{\sqrt2}\begin{pmatrix}
   \sin\phi+i\xi \cos\phi\cos\theta\\
  -\cos\phi+i\xi \sin\phi\cos\theta\\
  -i\xi \sin\theta
\end{pmatrix}.
\end{equation}
\end{widetext}
Therefore
\begin{equation*}
    \frac{\epsilon^x_{\bs q,\xi}\pm i\epsilon^y_{\bs q,\xi}}{\sqrt2}=\mp \frac{ie^{\pm i\phi_{\bs{q}}}}{2}\left(1\mp\xi\cos\theta_{\bs q}\right)
\end{equation*}
and the coupling constant acquires the following form:
\begin{align*}
    &\lambda^{\pm}_{mn}(\bs q, \xi)=\hbar\omega_H{\psi}^j_{\rm ph}(l)\sqrt{\frac{e^2}{\varepsilon\hbar\omega_{\bs q}}\frac{2\pi l_H^2}{S}}\sqrt{{\rm max}(m,n)}\times\nonumber\\
    &\times\left(\frac{|\bs q(l)|\mp\xi q_z(l)}{2|\bs q(l)|}\delta_{m,n+1}+
    \frac{|\bs q(l)|\pm\xi q_z(l)}{2|\bs q(l)|}\delta_{m,n-1}\right),
\end{align*}
where we omit the phase factor $\mp ie^{\pm i\phi_{\bs{q}}}$.

This expression implies that 2DEG can only emit left photons only with increase of electron Landau level, whereas, the emission of right photons requires a decrease of Landau level (the physical reason is the angular momentum conservation). Thus, due to energy conservation only left photons can scatter to the condensate mode from the pumping ones and we further set $\xi=-1$:
\begin{equation}
    \lambda^{\pm}_{mn}(\bs 0)=\hbar\omega_H\psi_{\rm cond}^j\sqrt{{\rm max}(m,n)\frac{e^2}{\varepsilon\hbar\omega_c}\frac{2\pi l_H^2}{S}}\delta_{m,n\pm 1}.
    \label{eq:lambda-cond}
\end{equation}

\section{The impact of magnetic field gradient}\label{app:gradient}

To increase particle flux from pumping modes to the condensate mode, one needs to involve a bunch of them. One way to achieve that is introducing magnetic field gradient. Yet, for moderate gradient, one may not change the form of the Hamiltonian, as we demonstrate below.

The non-uniform magnetic field $H_z=H(1+\kappa x/R)$ may be introduced via the following vector potential in symmetric gauge:
\begin{equation}
    \textbf{A}_H=H\left(-y,\kappa\frac{x^2}{2R},0\right),
\end{equation}
where  $R$ is the linear size of the resonator, $\kappa$ is the magnetic field change percentage.

The dimensionless version of the Schrödinger equation for an electron in a non-uniform magnetic field is
\begin{equation}
    \left[\left(-i\nabla_\zeta-\eta\right)^2+\left(-i\nabla_\eta+\frac{\kappa l_H}{2R}\zeta^2\right)^2\right]\psi(\zeta,\eta)=\epsilon\psi(\zeta,\eta)
\end{equation}
where $\zeta = x/l_H$ and $\eta = y/l_H$ are dimensionless coordinates. 
Typical resonator lengths we consider are of the order of several millimeters, while $l_H\!\sim$nm. Thus:
\begin{equation*}
    \frac{l_H}{2R}\in [10^{-6};10^{-5}].
\end{equation*}
Clearly, since the pumping range can not exceed the width of a single minimum of the loss profile (see Fig. \ref{fig:scheme}), $\kappa$ is at most of the order of unity. In our calculations $\zeta^2 \le 100$. Thus, the additional term may be safely neglected when calculating all the rates. All the functions of magnetic field should be treated as averaged over the sample:
\begin{align}
    f(H) &= \langle f(H)\rangle \equiv \int_{-L/2}^{L/2} f\left(H(1+\kappa x/L)\right)dx/L.
\end{align}

\section{Calculation details for condensate mode kinetics}\label{app:calculations}
Here we address all the kinetic processes to derive expressions for the rates  $R$ \eqref{eq:nu-pump}, $r$ \eqref{eq:nu-reverse}, $\varkappa_{\rm loss}$ \eqref{eq:spontaneous_loss} and $\varkappa_{\rm sat}$ \eqref{eq:nu-sat} introduced in the main text.

\subsection*{Electronic degrees of freedom}
In a real semiconductor heterostructure the 2DEG will be confined in several $N_{\rm layer}$ layers in the center of the cavity. Thus, 
\begin{equation*}
    \sum_{\bs p, j}\equiv\sum_{j=1}^{N_{\rm layer}}\int\frac{dp_xL_x}{2\pi\hbar}\sum_n
\end{equation*}
is the sum over the layers, electron momenta and Landau levels (denoted by $n$).
In a cavity with a profile of the type we show in Fig. \ref{fig:resonator_profile} with electron gas localized in the region of high amplification, the wavelength is by several orders of magnitude higher than the total width of 2DEG layers. Thus, one may neglect wavefunction variation over the width of the 2DEG layers and simplify:
\begin{equation}
    \sum_{j=1}^{N_{\rm layer}}\!\!f\big(|\psi_{\rm ph}^j|^2\big){\simeq}\!\!\sum_{j=1}^{N_{\rm layer}}\!\!f(|\psi_{\rm ph}|^2){=}N_{\rm layer}f(|\psi_{\rm ph}|^2).
    \label{eq:electron-center}
\end{equation}
Here $\psi_{\rm ph}$ is the wavefunction amplitude in the center of the cavity.

Momentum summation may be performed separately and results in degeneracy of each level:
\begin{equation}
    \int_{-p_{\rm max}/2}^{p_{\rm max}/2}\frac{dpL_x}{2\pi\hbar}=\frac{p_{\rm max}L_x}{2\pi\hbar}=\frac{S}{2\pi l_H^2}.
\end{equation}
Here $S=L_x\times L_y$ is the area of the layer, $p_{\rm max}=L_y\hbar/l_H^2$.

Altogether:
\begin{equation}
    \sum_{\bs p}f(n,\psi_{{\rm ph},j}^2)=\frac{S}{2\pi l_H^2}N_{\rm layer}\sum_n f(n,\psi_{\rm ph}^2).
\end{equation}
\subsection*{Photonic degrees of freedom}
The summation over the photonic variables is over the in-plane momentum $\bs q$ and number of modes $l$.
\begin{equation*}
    \sum_{\bs q}\equiv\sum_l\int\frac{d^2\bs q_\parallel S}{(2\pi\hbar)^2}.
\end{equation*}
The terms in the photonic sum always come with $\lambda^\pm_{mn}(\bs q)$, which depends on the photonic wavefunction amplitude $\psi_{\rm ph}(l)$ in the center of the cavity. When performing summation over the photonic modes, one should consider the high-Q mode separately, to which we assign $l=0$ and address to it further.

The summation over $l$ may be performed explicitly, assuming the number of modes in the sum $N_l\gg 1$:
\begin{align}
    \sum_l\psi_{\rm ph}^2(l)&\left(\frac{1\pm\cos\theta_l}{2}\right)^2\simeq\int_0^{q_{\rm cut}}\frac{2dq_z}{8\pi\hbar}\left(1\pm\frac{q_z}{q}\right)^2\nonumber\\
    &=\frac{\omega_{\rm cut}\sqrt\varepsilon}{4\pi \hbar c}\left(1\pm\frac{\omega_{\rm cut}}{\omega_q}+\frac{\omega_{\rm cut}^2}{3\omega_q^2}\!\right).
\end{align}
Here $\omega_{\rm cut}$ is the maximal frequency of pumping modes.

\subsection*{(i) Condensate replenishment from the pumping modes}\label{subapp:pumping rate}
Utilizing the expressions derived above, we end up with the following expressions:

\begin{align}
    R&{=}\omega_{\rm cut}\alpha N_{\rm layer}\frac{e^2\psi_{\rm cond}^2}{\sqrt\varepsilon\hbar\omega_c}\!\!\left(\frac{l_H\omega_H^2}{c(\omega_c{+}\omega_r)}\right)^{\!\!2}\!\!\!\left(1{-}\frac{\omega_{\rm cut}}{\omega_{\rm res}}{+}\frac{\omega_{\rm cut}^2}{3\omega_q^2}\!\right)\nonumber\\
    &\times\sum_{n,s}(n+1)(n+2)N^{\rm el}_{n,s}\left(1-N^{\rm el}_{n+2,s}\right),\\
    r&{=}\omega_{\rm cut}\alpha N_{\rm layer}\frac{e^2\psi_{\rm cond}^2}{\sqrt\varepsilon\hbar\omega_c}\!\!\left(\frac{l_H\omega_H^2}{c(\omega_c{+}\omega_r)}\right)^{\!\!2}\!\!\!\left(1{-}\frac{\omega_{\rm cut}}{\omega_{\rm res}}{+}\frac{\omega_{\rm cut}^2}{3\omega_q^2}\!\right)\nonumber\\
    &\times\sum_{n,s}(n+1)(n+2)N^{\rm el}_{n+2,s}\left(1-N^{\rm el}_{n,s}\right).
\end{align}
Here $\alpha=e^2/\hbar c$ is the fine-structure constant and $\omega_r=\omega_H+\omega_s$ resonance photon absorption frequency. 

\subsection*{(ii) Phonon-assisted photon absorption by Landau-quantized electrons}\label{subapp:loss rate}

\begin{widetext}
Let's consider the loss rate function. It has two terms with phonon absorption and emission:
\begin{align*}
    &\varkappa_{\rm loss}{=}\frac{SN_{\rm layer}}{2\pi l_H^2}\!\sum_n\!\frac{2\pi}{\hbar}\!\!\int\!\frac{d^3kV}{(2\pi\hbar)^3}|A_{n,m}^{\rm loss}|^2\!\!\left(\!\frac{\delta(\hbar\omega_c{-}\hbar\omega_s{-}(m{-}n)\hbar\omega_H{-}ku)}{1{-}\exp(-ku/T)}{+}\frac{\delta(\hbar\omega_c{-}\hbar\omega_s{-}(m{-}n)\hbar\omega_H{+}ku)}{\exp(ku/T){-}1}\right)\left(N^{\rm el}(E_n^+){-}N^{\rm el}(E_m^-)\right).
\end{align*}
We can simplify the expression taking into account the linear dependence of the $|A_{\rm loss}|^2$ on the phonon energy $ku$:
\begin{equation*}
    \int_0^\infty  d^3k\frac{ku\delta(\Delta+ku)}{1-e^{-ku/T}}=\int_{-\infty}^0d^3k\frac{ku\delta(\Delta-ku)}{e^{ku/T}-1}
\end{equation*}
The integrand is nonzero near the $ku\sim \hbar u/l_H\ll T$:
\begin{align}
    \frac{2\pi}{\hbar}\int_0^\pi\frac{2\pi d\cos\theta}{(2\pi\hbar u)^3}\int_{-\infty}^\infty\frac{(ku)^2d(ku)V}{e^{ku/T}-1}
    |A_{n,m}^{\rm loss}|^2 \delta(\Delta_m-ku)\simeq\frac{\Delta_mTV}{2\pi\hbar(\hbar u)^3}\int_0^\pi d\cos\theta|A_{nm}^{\rm loss}(\Delta_m,\theta)|^2,
    \label{eq:loss-delta}
\end{align}
The absorption amplitude has a form:
\begin{align*}
    A^{\rm loss}_{nm}&{=}\sqrt{\!\frac{D_e^2 ku}{\rho u^2 V}\frac{e^2\psi_{\rm cond}^2}{\varepsilon\hbar\omega_c}\frac{2\pi l_H^2}{S}}
    V_{00}\!\left(\frac{k_z^2l_z^2}{2\hbar^2}\!\right)\!\!\left.\left(\frac{\sqrt{n}V_{n{\shortminus} 1,m}\left(x\right)\omega_H}{E_n^++\hbar\omega_c-E_{n-1}^+}{-}\frac{\sqrt{m{+}1}V_{n,m{+}1}\left(x\right)\omega_H}{E_{m+1}^++\hbar\omega_c-E_{m}^-}\right)\right|_{x={k_{xy}^2l_H^2}/{2\hbar^2}}
\end{align*}
Due to the neglected nonlinearity, we can simplify the entire expression:
\begin{equation*}
    E_n^++\hbar\omega_c-E_{n-1}^+=E_{m+1}^++\hbar\omega_c-E_{m}^-=\hbar\omega_c+\hbar\omega_r.
\end{equation*}
The loss amplitude may be simplified with the use of the properties of the Laguerre's polynomials and \eqref{eq:overlap_integral}: 
\begin{align*}
    A^{\rm loss}_{nm}&{=}\sqrt{\!\frac{D_e^2 ku}{\rho u^2 V}\frac{e^2\psi_{\rm cond}^2}{\varepsilon\hbar\omega_c}\frac{2\pi l_H^2}{S}}\frac{\omega_H}{\omega_r{+}\omega_c}
    V_{00}\!\left(\frac{k_z^2l_z^2}{2\hbar^2}\!\right)\!\!\left.\left({\sqrt{n}V_{n{\shortminus} 1,m}\left(x\right)}{-}{\sqrt{m{+}1}V_{n,m{+}1}\left(x\right)}\right)\right|_{x={k_{xy}^2l_H^2}/{2\hbar^2}}\\
    &\qquad\qquad\quad={-}\sqrt{\frac{D_e^2 ku}{2\rho u^2 V}\frac{e^2\psi_{\rm cond}^2}{\varepsilon\hbar\omega_c}\frac{2\pi l_H^2}{S}}\frac{\omega_H}{\omega_r{+}\omega_c}V_{00}\!\left(\!\frac{k_z^2l_z^2}{2\hbar^2}\!\right)\!\!\left.\sqrt{x}V_{nm}\left(x\right)\right|_{x={k_{xy}^2l_H^2}/{2\hbar^2}}\!,
\end{align*}
where $k<0$/$k>0$ stands for phonon absorption/emission, $\Delta_m=\hbar\omega_c-\hbar\omega_s-m\hbar\omega_H$ is the photon energy offset from resonant absorption frequency. 

Altogether:
\begin{align}
    \varkappa_{\rm loss}(\hbar\omega_c)&=\omega_H\frac{D_e^2\hbar\omega_HT}{2\pi\rho u^2(\hbar u)^3}\frac{e^2\psi_{\rm cond}^2N_{\rm layer}}{\varepsilon\hbar\omega_c}\sum_{n,m}\left(\frac{\Delta_m}{\hbar\omega_r+\hbar\omega_c}\right)^2\left(N^{\rm el}(E_n^+){-}N^{\rm el}(E_m^-)\right)\times\\
    &\left.\times\int_0^\pi\frac{d\cos\theta}{4}\left(\frac{\Delta_m l_H\sin\theta}{\hbar u}\right)^2V_{00}^2\left(\frac{\Delta_m^2l_z^2}{2\hbar^2 u^2}\cos^2\theta\right)V_{nm}^2\left(\frac{\Delta_m^2l_H^2}{2\hbar^2 u^2}\sin^2\theta\right)\right|_{\Delta_m=\hbar\omega_c-\hbar\omega_s-(m-n)\hbar\omega_H}
\end{align}
From this expression one may infer the width of the loss peak:
\begin{equation*}
    \int d\cos\theta \sin^{m-n+2}(\theta)\exp\left(-\frac{\Delta_m^2}{2\hbar^2u^2}(l_H^2\sin^2\theta+l_z^2\cos^2\theta)\right).
\end{equation*}
\end{widetext}

\subsection*{(iii) Interaction with satellite modes}\label{subapp:sattelite rate}

In the same fashion as for the $R$ and $r$, one may consider the interaction with the satellite modes:
\begin{align}
    \varkappa_{\rm sat}&=\frac{2\pi}{\hbar}\!\!\!\sum_{\substack{{\rm electron}\\{\rm photon}}}\!\!\!\left(\frac{\lambda_{n-s,n}^-(q_{\rm sat},l)\lambda^{c+}_{n,n-s}}{E^s_{n}+\hbar\omega_c-E_{n-s}^{-s}}\right)^2\nonumber\\
    &\times\delta(\hbar\omega_{\rm sat}-\hbar\omega_c) N^{\rm el}_T(n^s)\left(1-N^{\rm el}_T(n^s)\right).
\end{align}
What is different is only the sum over Landau levels:
\begin{align}
    \varkappa_{\rm sat}&=\omega_{\rm cut}\alpha N_{\rm layer}\frac{e^2\psi_{\rm cond}^2}{\sqrt\varepsilon\hbar\omega_c}\!\!\left(\frac{l_H\omega_H^2}{c(\omega_c{+}\omega_r)}\right)^{\!\!2}\!\!\!\left(1{+}\frac{\omega_{\rm cut}}{\omega_{c}}{+}\frac{\omega_{\rm cut}^2}{3\omega_{c}^2}\right)\nonumber\\
    &\times\sum_{n,s}\left({\rm max}(n,n-s)\right)^2N^{\rm el}_T(n^s)\left(1-N^{\rm el}_T(n^s)\right).
\end{align}

\section{Electron level broadening}\label{app:broadening}

The key component of the proposed condensation mechanism is equal spacing between the electron Landau levels. In general, in real-life materials due to multiple unaccounted scattering channels (phonons, disorder, \textit{etc.}), electron mobility is finite, which results in some broadening of electron levels and increase of losses. To estimate the significance of this effect, we consider a Gaussian broadened spectral function of a Landau electron:
\begin{align}
    \mathcal A (n, \omega)=\frac{\sqrt{2\pi}}{\Gamma}\exp\left(-\frac{(\omega - n\omega_H)^2}{2\Gamma^2}\right).
\end{align}
Here $\tau = {\hbar}/{\Gamma}$  is the mean free time for an electron in 2DEG.

The mobility is given as $\mu = {e\tau}/{m^*}$, thus:
\begin{align}
    \Gamma = \frac{e\hbar}{m^*\mu}.
\end{align}

We utilize this expression to replace the delta-function in \eqref{eq:loss-delta} to account for broadening. The resultant broadening is shown in Fig. \ref{fig:leak-map}.

\section{Conditions for neglecting the nonlinearity}\label{app:nonlinearity}

In the main text we assumed that kinetic rates are independent of condensate density. By this we neglected the influence of highly populated photonic modes on electronic states. Here we estimate the validity of this assumption.

\begin{figure}
    \begin{center}
    \includegraphics[width=\linewidth]{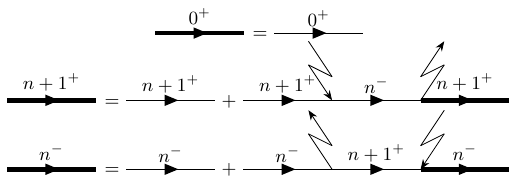}
    \end{center}
    \caption{The Dyson equation for electrons in photon condensate medium. Solid lines denote electrons, zigzags stand for condensate photon (following Beliaev's conventions).}
    \label{fig:dyson}
\end{figure}

The electron Green's function in presence of condensate is dressed as shown by the Dyson equations presented in Fig. \ref{fig:dyson}.

With neglection of condensate fluctuations (as in Beliaev technique \cite{Beliaev58}), the corresponding equation system is solved straightforwardly:
\begin{align}
    E_{n}^{s,\rm int}-E_{n}^{s}&\equiv \Sigma^s_n,\\
    \Sigma_0^+ &= 0.
\end{align}
Thus, interaction with condensate results in shift electron levels by:
\begin{align}
    \Sigma_{n+1}^\pm &= \pm\frac{\hbar\omega_r}{2}\left(\sqrt{1+\left|\frac{2\lambda_{n+1,n}(\bs 0)}{\hbar\omega_r}\right|^2N_0}-1\right).
\end{align}
In order for these shifts to be negligible, they should be small compared to the distance between loss peaks:
\begin{align}
    \frac{\delta E}{\hbar\omega_H}\ll1, 
\end{align}
which implies small value of $n$. We can expand the energy shifting to the first order of $n$:
\begin{equation}
    {\Sigma_n(n_0)}\simeq \frac{\hbar\omega_r}{4}\left|\frac{2\lambda(\bs0)}{\hbar\omega_r}\right|^2N_0\simeq2\pi n\frac{e^2\psi_c^2}{\varepsilon}n_0l_H^2.
\end{equation}
The lowest energy scale of the system is set by the phonon energy $\hbar u/l_H$. So, the critical density is the one, for which $\Sigma_n(n_0^{\rm crit})\sim\hbar u/l_H$:
\begin{equation}
    n_0^{\rm crit}(n)\sim\frac{\varepsilon \hbar u}{2\pi n e^2\psi_c^2l_H^3}\sim\frac{m_e^*u}{\alpha\hbar l_H}.
\end{equation}
The Landau level  is about $10^1-10^2$, so the critical photon density is about $10^{12}$ cm$^{-2}$.
We can see that the limit of the black body radiation flux imposes much stronger constraint on the final condensate state:
\begin{align}
    W{\simeq}\frac{cS_{\rm res}}{\sqrt\varepsilon}\int_{\hbar\omega_{\rm res}^{\rm min}(q)}^{\hbar\omega_{\rm res}^{\rm max}(q)}\!\frac{d^3q}{(2\pi\hbar)^3}N_{BE}\left({T_{\rm BB}}\right)\simeq\nonumber\\
    \simeq
    \frac{\varepsilon\omega_{\rm res}^2\Delta\omega_{\rm res}N_{\rm BE}(T_{\rm BB})}{2\pi^2c^2}<\varkappa \frac{\omega_H\Delta\omega}{c^2} Q
\end{align}
The photon density by blackbody pumping has an upper limit:
\begin{equation}
    n_0^{\rm BB}<\frac{W}{\varkappa_0+\varkappa}< Q\frac{\omega_H\delta\omega}{c^2}\ll\frac{m_e^*u}{\alpha\hbar l_H}\sim n_0^{\rm crit}.
\end{equation}
We can compare the critical density and photon density achieved by blackbody pumping: $n_0^{\rm BB}\ll n_0^{\rm crit}$. Thus, nonlinearities are safely negligible for our system.

\bibliography{apssamp}

\end{document}